\documentclass[12pt]{article}%
\pdfoutput=1

\usepackage{amsmath,amssymb,amsfonts}
\usepackage{color}
\definecolor{darkblue}{rgb}{0.1,0.1,.7}
\usepackage[colorlinks, linkcolor=darkblue, citecolor=darkblue, urlcolor=darkblue, linktocpage]{hyperref} 
\usepackage[square, comma, compress,numbers]{natbib}
\usepackage[]{amsmath}
\usepackage[]{graphicx}
\usepackage[]{latexsym}
\usepackage{geometry}
\usepackage{amscd}
\usepackage[all,cmtip]{xy}
\usepackage{mathrsfs}
\usepackage[margin=10pt,font=small,labelfont=bf]{caption}
\def\nn{\nonumber} 
\renewcommand{\epsilon}{\varepsilon}
\def\eps{\epsilon}
\renewcommand{\mathcal}{\mathscr}

\newcommand{\reef}[1]{(\ref{#1})}
\newcommand{\hhref}[1]{\href{http://arxiv.org/abs/#1}{arXiv:#1}}
\def\beq{\begin{equation}} 
\def\eeq{\end{equation}} 
\def\del{\partial}
\numberwithin{equation}{section}

\begin{document}

\begin{titlepage}
\begin{flushright}
LPTENS-11/41
\end{flushright}
\vskip 1.0cm
\begin{center}
{\Large \bf Exploring $T$ and $S$ parameters\\[3mm] 
in Vector Meson Dominance Models\\[5mm] 
of Strong Electroweak Symmetry Breaking}
\vskip 1.0cm
{\large Axel Orgogozo$^{a}$ and Slava Rychkov$^{a,b}$} \\[0.7cm]
{\it $^a $
Laboratoire de Physique Th\'{e}orique, \'{E}cole Normale Sup\'{e}rieure, Paris, France\\[5mm]
$^b$ Facult\'{e} de Physique, Universit\'{e} Pierre et Marie Curie, Paris, France}
\end{center}
\vskip 1.0cm
\begin{abstract}
We revisit the electroweak precision tests for Higgsless models of strong EWSB. We use the Vector Meson Dominance approach and express $S$ and $T$ via couplings characterizing vector and axial spin-1 resonances of the strong sector. These couplings are constrained by the elastic unitarity and by requiring a good UV behavior of various formfactors.
We pay particular attention to the one-loop contribution of resonances to $T$ (beyond the chiral log), and to how it can improve the fit. We also make contact with the recent studies of Conformal Technicolor. We explain why the second Weinberg sum rule never converges in these models, and formulate a condition necessary for preserving the custodial symmetry in the IR.

\end{abstract}
\vskip 1cm \hspace{0.7cm} November 2011
\end{titlepage}

\newpage

\tableofcontents

\newpage

\section{Introduction}
\label{sec:intro}

The purpose of this paper is to revisit the problem of electroweak precision tests in strong electroweak symmetry breaking (EWSB) models. 

Remember that EWSB scenarios can be roughly classified into models with or without a Higgs boson, by which we mean one or more scalar particle with a significant coupling to $WW$, so that it plays a dominant role in unitarizing the longitudinal $WW$ scattering.

Models with a Higgs boson are often called \emph{weakly coupled}, because they can be extrapolated up to energies $\Lambda_{UV}\gg\Lambda_{EW}\sim4\pi v$ ($v=246$ GeV is the EWSB scale). For example, for low energy SUSY models one can have $\Lambda_{UV} \sim \Lambda_{GUT}$, while for models of composite pseudo-goldstone Higgs boson one typically expects ${\Lambda_{UV}}/{\Lambda_{EW}}= O(3\div5)$ \cite{SILH,extended}.

On the other hand, \emph{strongly coupled} models, in which $\Lambda_{UV}\sim\Lambda_{EW}$, are usually assumed to give rise not to a Higgs boson, but to a sequence of vector resonances $V$ with masses $M_V= O(\Lambda_{EW})$. As is well known, vector exchanges are also able to unitarize $W_L W_L$ scattering \cite{Bagger,Chivukula-unit,us}. This unitarization remains imperfect, since unitarity violations are postponed to a scale which is only a few times higher than $M_V$. Still, this possibility is very interesting. The unitarization condition fixes the $V$ coupling to $W_L W_L$ in terms of its mass, and gives rise to a predictive framework. In particular, the $V\rightarrow W_L W_L$ decay width can be calculated and is found relatively narrow ($\Gamma_V/\Gamma_H=O(0.1)$ relative to the Higgs boson width for the same mass).   

Historically, the first strongly coupled EWSB model was Technicolor \cite{WS}. In Technicolor, EWSB sector originates from an asymptotically free gauge theory in the UV, with a rapid confinement and chiral symmetry breaking in the IR, very much like in QCD. 

However, other types of UV dynamics may also lead to strong EWSB. One possibility is that the UV theory stays close to a strongly interacting conformal fixed point over a wide range of energies, losing any connection to the weakly coupled description in the deep UV (if there was any)  \cite{Holdom,Luty}. In this case, thinking in terms of gauge degrees of freedom may be misleading. One may try instead to use the language of Conformal Field Theory \cite{R,V,DDV}. Conformal language is also indispensable in the context of strong EWSB models in warped extra dimensions \cite{RS}.

The defining feature of strong EWSB models is thus the presence of strong interactions and composite resonances at $\Lambda_{EW}$, and not the nature of fundamental degrees of freedom out of which these resonances are built. It is then interesting to know if the electroweak precision observables can be computed in terms of resonance masses and couplings. This is the question that we attempt to answer here.

The paper is structured as follows. We begin in section \ref{sec:brief} with a brief reminder of the electroweak precision test (EWPT) formalism. 

In section \ref{sec:VMD} we introduce an effective lagrangian for the strong EWSB sector. In addition to the usual goldstones, it contains the spin-1 vector and axial resonances. They transform nonlinearly under the $SU(2)_L\times SU(2)_R$ global symmetry, as is appropriate for the composite particles. Their couplings are fixed by the assumption that the UV behavior of various amplitudes ($\pi\pi$ scattering, pion formfactor) is improved by tree-level resonance exchanges. This framework, called Vector Meson Dominance (VMD), is experimentally known to work well in QCD \cite{Ecker0}, and we adopt it here.

In section \ref{sec:S} we discuss the $S$ parameter. In sections \ref{sec:PT}, \ref{sec:SVMD} we review the Peskin-Takeuchi dispersion relation and use it to compute $S$ in our model. Most of the discussion here is not new. Our explanation why the chiral log contribution is expected to be cutoff at $\Lambda\sim M_V$ in VMD models may be not without interest. 

In section \ref{sec:UVtail} we discuss the Weinberg sum rules. We explain why the second sum rule cannot be expected to converge in models of strong EWSB which resolve the flavor problem via large Higgs anomalous dimension (Walking or Conformal Technicolor). This fact has been suspected before, but can now be shown rigorously, using recent results about the UV conformal structure of such models.

In section \ref{sec:T} we discuss the $T$ parameter. In section \ref{sec:gold} we justify the prescription relating $T$ to the pion wavefunction renormalization in the Landau gauge. In section \ref{sec:TUV} we address (the absence of) the UV sensitivity. We point out that custodial symmetry in the IR is not an automatic consequence of the UV fixed point custodial invariance: it could be broken by scalars which are singlets of $SU(2)_W\times U(1)_Y$ but not of $SU(2)_L\times SU(2)_R$. Demanding that all such dangerous scalars be irrelevant is an extra condition on viable Conformal Technicolor models.

In section \ref{sec:TVMD} we proceed to compute the one-loop $O(g'^2)$ contribution to the $T$ parameter from goldstones and resonances, in our VMD model. Our discussion here completes and generalizes that of Ref.~\cite{us}. In particular, we consider the full list of cubic vector-axial-goldstone couplings relevant for this computation. Two of these can be fixed by demanding that they regulate the formfactors of spin-1 resonances, thus reducing the UV sensitivity of $T$ from quadratic to logarithmic. The third coupling is a free parameter.

Finally, in section \ref{sec:vs} we present numerical results showing how our VMD model compares with the electroweak precision data, and identify preferred regions of the parameter space. We conclude in section \ref{sec:concl}. Appendices \ref{sec:eps13exp},\ref{sec:piAV},\ref{sec:Tbulky} collect technical details related to the electroweak fit and to the $T$ parameter computation.

\section{Brief Reminder of Electroweak Precision Analysis}
\label{sec:brief}

Corrections to the electroweak precision observables can occur via the weak boson self-energies (\emph{oblique}, or \emph{universal} corrections), or in fermion-weak boson vertices. In this paper we will be concerned only with the oblique corrections. In general, the vertex corrections are more model-dependent, since they require a discussion of how fermions couple to the EWSB sector. Sometimes large vertex corrections are invoked to improve the electroweak fit \cite{Cacciapaglia:2004rb}, but that's not the road we would like to explore here. Rather, we would like to see if a satisfactory fit can be obtained in the absence of significant vertex corrections.

Let us call a EWSB model \emph{heavy}, if all new particles/resonances have masses $ \gg M_Z$ (including the Standard Model Higgs boson if it exists). In this paper we will be mostly dealing with such models. 

In heavy EWSB models, electroweak precision tests are convenient to perform in terms of the three $\epsilon$ parameters \cite{barbieri:epsilons}.
On the one hand, the $\epsilon$'s are linearly related to the precision observables $\Delta \rho$, $\Delta k$, and $\Delta r_\text{w}$ by universal coefficients dependent only on $s_W$, the sine of the Weinberg angle. The latter observables describe genuine electroweak corrections beyond the running of $\alpha_{\rm EM}$. Here we will need only $\epsilon_1$, $\epsilon_3$, determined experimentally as\footnote{This determination assumes that the $\epsilon_2$ parameter does not deviate significantly from its SM value. This is justifiable in heavy EWSB models, where one expects $\Delta \eps_2\ll\Delta\eps_1$. Analogously, the variations of the $Y$ and $W$ parameters, constrained by LEP2 data \cite{LEP2}, are also expected to be negligible in our models.} (Appendix~\ref{sec:eps13exp})
\begin{equation}
\epsilon_1 =(6.0\pm0.7)\,10^{-3}\,,\qquad \epsilon_3 =(5.7 \pm 0.8)\,10^{-3}\,,
\label{eq:eps13exp}
\end{equation}
with $85 \%$ correlation. On the other hand, the same $\epsilon$'s can be computed in terms of the gauge boson self-energies \cite{barbieri:epsilons}
\begin{equation} 
\begin{split}
\epsilon_1 &= e_1 + \delta\epsilon_1,\qquad e_1 = \frac{\Pi_{33}(0)-\Pi_{+-}(0)}{M_W^2}\,,   \\
\epsilon_3 &= e_3 + \delta \epsilon_3,\qquad e_3 = \frac{c_W}{s_W}\Pi'_{3B}(0)\,,\label{eq:eps13}
\end{split}
\end{equation}
where $\Pi_{ij}$ are the formfactors appearing in the gauge boson vacuum polarization amplitudes:
\begin{equation}
\Pi_{ij,\mu\nu}(q)=-i\eta_{\mu\nu} \Pi_{ij}(q^2)+ q_\mu q_\nu \text{ terms}\,.
\end{equation}

The terms $\delta \epsilon_{1,3}$ in (\ref{eq:eps13}) involve combinations of gauge boson self-energies which are higher order in the $q^2$ expansion. We will not need their precise form, which can be found in \cite{barbieri:epsilons}. In heavy models, $\delta \epsilon_{1,3}$ are dominated by $W,Z$ loops and are not sensitive to heavy new particles. Still, it is useful to remember about their presence in (\ref{eq:eps13}). In particular, this helps to clarify issues related to the gauge invariance of $\epsilon_{1,3}$. 

\begin{figure}[htbp]
\begin{center}
\includegraphics[scale=0.6]{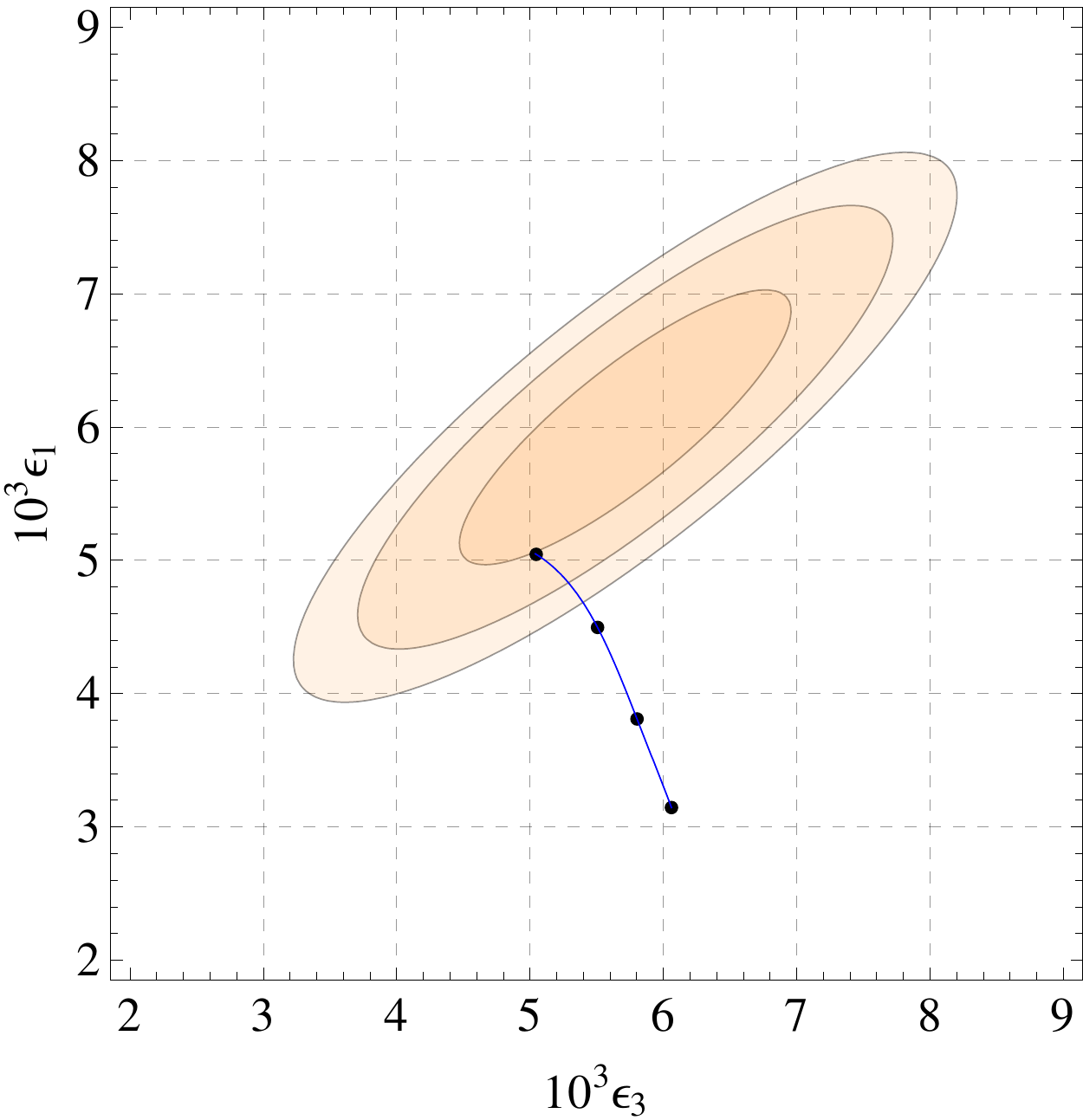}
\caption{Experimental contours (68,95 and 99\% CL) for $\epsilon_{1,3}$, as well as their dependence on the Higgs mass in the SM ($M_H=115, 200, 500, 1000$ GeV). The top mass is set at the current central value $M_t=173.3$ GeV.}
\label{fig:e13SM}
\end{center}
\end{figure}

For the Standard Model (SM), comparison of theory and experiment points to a relatively light Higgs boson. This well known fact is demonstrated in Fig.\ \ref{fig:e13SM}.
For a heavy Higgs, the $\epsilon_1$, $\epsilon_3$ have a logarithmic dependence on $M_H$:
\begin{equation}
\epsilon_1^{\rm SM} \approx -\frac{3 g'^2}{32 \pi^2}\log \frac{M_H}{M_Z}+\text{const},\qquad
\epsilon_3^{\rm SM} \approx \frac{g^2}{96 \pi^2}\log \frac{M_H}{M_Z}+\text{const}',
\label{eq:asympteps}
\end{equation}
up to $O(M_Z^2/M_H^2)$ corrections. These logarithms come from $e_1$ and $e_3$, while $\delta \eps_{1,3}$ go to a constant in the heavy Higgs limit.

The curve in Fig.~\ref{fig:e13SM} has been  traced using the full one-loop formulas from \cite{Novikov}; the fact that it deviates from a straight line in the light Higgs region shows that the asymptotic expressions (\ref{eq:asympteps}) are not accurate there.

For any other model, electroweak precision analysis is simplified by focussing on the deviations of $\epsilon_{1,3}$ from their values in the SM for a reference value of the Higgs mass. These deviations are known as the $T$ and $S$ parameters\footnote{$\hat{T}=\alpha T$, $\hat{S}=\frac{\alpha}{4 s_W^2}S$ compared to the normalization of \cite{Peskin}.}:
\begin{equation}
\begin{split}
\hat{T} &=\epsilon_1-\epsilon_1^{\rm SM} \approx e_1-e_1^{\rm SM}\,, \\
\hat{S} &=\epsilon_3-\epsilon_3^{\rm SM} \approx e_3-e_3^{\rm SM}\,, 
\end{split}
\end{equation}
where the approximations are valid if the model is heavy and $M_H^{\text{ref}}\gg M_Z$.

\section{Strong EWSB and Vector Meson Dominance}
\label{sec:VMD}

We assume as usual that the strong EWSB sector has a custodial $SU(2)_L\times SU(2)_R$ global symmetry.
The corresponding symmetry currents are $J_L^{a\mu}$ and $J_R^{a\mu}$, and we will also use their vector and axial combinations $J_{V,A}^\mu=J_L^{\mu}\pm J_R^{\mu}$. For simplicity, we assume that the strong sector preserves parity, under which $J_L^{\mu}\leftrightarrow J_R^{\mu}$,  $J_A^{\mu} \to -J_A^{\mu}$.\footnote{See \cite{Falkowski} for a recent discussion of strong EWSB without parity.} 

The SM gauge group partly gauges the global symmetry of the EWSB sector. In the standard convention, $SU(2)_W$ gauges $J_L^{\mu}$, while $U(1)_Y$ gauges $J_R^{3\mu}$.

Finally, by assumption, the global symmetry of the EWSB sector is broken spontaneously to the diagonal subgroup $SU(2)_{L+R}$. This triggers EWSB as it gives rise to three goldstone bosons $\pi^a$, eventually eaten by the $W$ and $Z$.\footnote{Notice that larger global symmetries of the strong sector, containing $SU(2)_L\times SU(2)_R$ as a subgroup, can be also considered. Such non-minimal models will contain extra (pseudo)-goldstone bosons in addition to the longitudinal $W$ and $Z$. In particular, a light composite Higgs boson can emerge as one of these pseudo-goldstones; see \cite{Ratt-recent} for a recent discussion.}

An effective low-energy description of this minimal EWSB sector is provided by the chiral Lagrangian:
\begin{equation}
\mathcal{L}=\frac{v^{2}}{4}\langle D_{\mu}U(D^{\mu}%
U)^{\dagger}\rangle~, \label{eq:L2}%
\end{equation}
where
\begin{align}
&  U=e^{i2\hat{\pi}/v},\qquad\hat{\pi}=T^{a}\pi^{a}=\frac{1}{\sqrt{2}}\left[
\begin{array}
[c]{cc}%
\frac{\pi^{0}}{\sqrt{2}} & \pi^{+}\nonumber\\
\pi^{-} & -\frac{\pi^{0}}{\sqrt{2}}%
\end{array}
\right]  ~,\qquad T^{a}=\frac{1}{2}\sigma^{a},\nonumber\\
&  D_{\mu}U=\partial_{\mu}U-i\hat{B}_{\mu}U+iU\hat{W}_{\mu}~,\qquad\hat
{W}_{\mu}=gT^{a}W_{\mu}^{a}~,\qquad\hat{B}_{\mu}=g^{\prime}T^{3}B_{\mu}~,
\label{def}%
\end{align}
and $\langle\rangle$ denotes the trace of a $2\times2$ matrix. The invariant
kinetic and mass terms for the SM fermions and gauge boson are left understood. Under
$SU(2)_{L}\times SU(2)_{R}$,
\begin{equation}
U\rightarrow g_{R}Ug_{L}^{\dagger}~.\nonumber
\end{equation}

The chiral Lagrangian is applicable at energies below any resonances. Such resonances can however be added to the model. For example, Higgs boson $h$ can be introduced via \cite{Bagger}
\begin{equation}
\mathcal{L}=\frac{1}{2}(\partial_{\mu}h)^2-\frac{M_H^2}{2}h^2+\frac12 hv\left\langle (D_\mu U)(D^\mu U)^\dagger\right\rangle\,.
\end{equation}
The last term describes the $h\pi \pi$ coupling, crucial for the unitarization of $\pi \pi$ (or $W_L W_L$) scattering. It also dominates the heavy Higgs decay width:
\begin{equation}
\Gamma_H\approx\Gamma(H\rightarrow \pi\pi)=\frac{M_H^3}{8\pi v^2} \ \ \ (M_H \gg M_Z)\,.
\label{eq:wH}
\end{equation}

In this paper, we are interested in models of strong EWSB, which are usually not expected to contain Higgs-like scalars. Instead, they may contain heavy spin-1 resonances (analogs of the $\rho$ and $a$ mesons in QCD).
From QCD literature \cite{Ecker}, there are several known, equivalent methods to couple such resonances to the chiral Lagrangian.
 
Here we will use a description by means of antisymmetric tensors transforming in the adjoint representation of $SU(2)_{L+R}$ (and nonlinearly under 
$SU(2)_{L}\times SU(2)_{R}$). Using antisymmetric tensors is a matter of convenience: up to field redefinitions, addition of local terms, and appropriate matching conditions for the coupling constants, equivalent results could be obtained with other formalisms. 
However, the issue of convenience should not be underestimated. There are at least three reasons why the antisymmetric tensor formalism looks better than the alternatives\footnote{A disadvantage of the antisymmetric tensor formalism is that the current automated collider physics tools do not deal with antisymmetric tensors. For this reason collider studies of Higgsless models are usually performed in the other two formalisms \cite{Bar1,Falkowski}.}  \cite{Ecker}:
\begin{itemize}
\item
It allows a unified treatment of states of positive and negative parity. The $J^P=1^+$ states could be introduced as gauge fields of $SU(2)_{L+R}$, in an approach known as ``hidden local symmetry"; however there is no simple natural way to do the same for the axials.\footnote{Axial spin-1 states can be introduced via the 4-site model \cite{4-site}. An equivalent construction which allows to introduce both vectors and axials as gauge fields was discussed recently in \cite{Redi}.}
\item
If vector fields in the adjoint of $SU(2)_{L+R}$ are used instead of antisymmetric tensors, extra local terms have to be added to the lagrangian to soften the UV behavior of various formfactors. In the antisymmetric tensor formalism these terms are accounted for automatically, due to a local term in the propagator (see below).
\item
Finally, using antisymmetric tensors avoids mixing of the spin-1 axial fields with the derivatives of the
goldstones.
\end{itemize}
One more reason will be mentioned in note \ref{note:4th}; see section \ref{sec:S} below. For a fair comparison, we must admit that the ``hidden local symmetry'' formalism has an advantage in that it allows to reach a weak coupling limit.\footnote{This was exploited recently in \cite{Panico:2011pw,Contino}.} However, this advantage would not be decisive for us, since we will be dealing here with theories which are on the verge of becoming non-perturbative.

We thus consider two sets of vector states of opposite parity, $V^{\mu\nu}$ and $A^{\mu\nu}$, both transforming in the adjoint representation of
$SU(2)_{L+R}$:
\begin{equation}
R^{\mu\nu}\rightarrow hR^{\mu\nu}h^{\dagger}~,\qquad R^{\mu\nu}=V^{\mu\nu
},\ A^{\mu\nu}~. \label{eq:Rtr}%
\end{equation}
We will be mostly considering the minimal situation when there is just one prominent resonance of each type, but it's trivial to generalize to several $V$'s and $A$'s. To describe the transformation properties of these fields under the full $SU(2)_{L}\times
SU(2)_{R}$, one introduces the little matrix $u$ \cite{coleman} via%
\begin{equation}
U=u^{2}~.\nonumber
\end{equation}
This matrix parametrizes the $SU(2)_{L}\times SU(2)_{R}/SU(2)_{L+R}$ coset and
transforms as%
\begin{equation}
u\rightarrow g_{R}uh^{\dagger}=hug_{L}^{\dagger}~,\nonumber
\end{equation}
where $h=h(u,g_{L},g_{R})$ is uniquely determined by this equation. The
general transformation of $R^{\mu\nu}$ is then given by the same Eq.\ (\ref{eq:Rtr}) with $h$ so defined. For $g_{L}=g_{R}$ we have $h=g_{L}=g_{R}$
independent of $u$, and we recover the linear $SU(2)_{L+R}$ transformation. This is the usual theory of nonlinear realizations \cite{coleman}.

The kinetic Lagrangian for heavy spin-1 fields has the form~\cite{Ecker0,Ecker}
\begin{equation}
\mathcal{L}=-\frac{1}{2}\langle\nabla_{\mu}%
R^{\mu\nu}\nabla^{\sigma}R_{\sigma\nu}\rangle+\frac{1}{4}M_{R}^{2}\langle
R^{\mu\nu}R_{\mu\nu}\rangle~,
\label{eq:lagr}
\end{equation}
where the covariant derivative
\begin{equation}
\nabla_{\mu}R=\partial_{\mu}R+[\Gamma_{\mu},R],\qquad\Gamma_{\mu}=\frac{1}%
{2}\left[  u^{\dagger}(\partial_{\mu}-i\hat{B}_{\mu})u+u(\partial_{\mu}%
-i\hat{W}_{\mu})u^{\dagger}\right]  ,\quad\Gamma_{\mu}^{\dagger}=-\Gamma_{\mu
},
\end{equation}
ensures that $\nabla_{\mu}R$ transforms as $R$ under the global $SU(2)_{L}%
\times SU(2)_{R}$ and under the SM gauge group.

We then have the following two-derivative $SU(2)_{L}\times SU(2)_{R}$ and parity invariant Lagrangian
describing the couplings of spin-1 fields to the
goldstones and SM gauge fields:
\begin{equation}
\mathcal{L}_{\text{int}}=\frac{i G_V}{2\sqrt{2}}\langle V^{\mu\nu}[u_{\mu
},u_{\nu}]\rangle+\frac{F_V}{2\sqrt{2}}\langle V^{\mu\nu}(u\hat{W}^{\mu\nu
}u^{\dagger}+u^{\dagger}\hat{B}^{\mu\nu}u)\rangle+\frac{F_A}{2\sqrt{2}}%
\langle A^{\mu\nu}(u\hat{W}^{\mu\nu}u^{\dagger}-u^{\dagger}\hat{B}%
^{\mu\nu}u)\rangle~, \label{eq:LV}%
\end{equation}
where
\begin{equation}
u_{\mu}=iu^{\dagger}D_{\mu}Uu^{\dagger}=u_{\mu}^{\dagger},~\qquad u_{\mu
}\rightarrow hu_{\mu}h^{\dagger}~.
\end{equation}
Parameters $G_{V},F_{V,A}$ have dimension of mass and by
Naive Dimensional Analysis (NDA) we expect them to be $O(v)$. 

Phenomenology taking into account tree-level exchanges of $V$ and $A$ resonances in the goldstone scattering amplitudes and in their coupling to the SM gauge fields is known as Vector Meson Dominance (VMD) \cite{Ecker0,Ecker}. We will now review a few of its features important for the EWSB, following \cite{us}.

Parameter $G_V$ measures the strength of the $V\pi\pi$ coupling. The basic elastic $\pi\pi$ scattering amplitude has the form \cite{Bagger,us}
\beq
\mathcal{A}= \frac
{s}{v^{2}} - \frac{G^{2}_{V}}{v^{4}} \left[  3 s + M_{V}^{2} \left(  \frac{s-u
}{t-M_{V}^{2}} + \frac{s-t}{u-M_{V}^{2}}\right)  \right]  ~, \label{eq:App}%
\eeq
where the first term comes from the chiral Lagrangian, while the second one is due to the $V$ exchange. It is derived by using the heavy vector propagator \cite{Ecker0,Ecker} 
\begin{gather}
\Delta^{ab}_{\mu\nu,\rho\sigma}\equiv \langle R^a_{\mu\nu}(k) R^b_{\rho\sigma}(-k)\rangle
=\delta^{ab}\Delta_{\mu\nu,\rho\sigma}\qquad (R_{\mu\nu}=R^a_{\mu\nu}T^a) \nn \\
\Delta_{\mu\nu,\rho\sigma}=-\frac{2i}{M_R^2}\left [
\frac{g_{\mu\rho} k_\nu k_\sigma-g_{\mu\sigma} k_\nu k_\rho-(\mu\leftrightarrow \nu)}
{k^{2}-M_R^{2}}
+(g_{\mu\sigma}g_{\nu\rho}-g_{\mu\rho}g_{\nu\sigma})\right],
\label{eq:Rprop}%
\end{gather}%
Notice the second, contact, term. This term appears automatically when deriving the propagator from the Lagrangian \reef{eq:lagr}. Its presence is important for ensuring the good UV behavior of the amplitude \reef{eq:App}. When using the description by nonlinearly realized \emph{vectors}, the contact term is absent and the amplitude grows as $s^2$. This growth should then be cancelled by adding a $\langle [u_\mu,u_\nu]^2\rangle$ term to the Lagrangian.
When using the antisymmetric tensor description, as we are doing here, one assumes that there is no extra $\langle [u_\mu,u_\nu]^2\rangle$ term with a significant coefficient, so that Eq.~\reef{eq:App} is a good approximation for the $\pi\pi$ scattering amplitude up to energies $\sim 2 M_V$. 

We now discuss consistency of Eq.~\reef{eq:App} with elastic unitarity. As is well known, the first term in the amplitude (pure chiral Lagrangian contribution) would violate unitarity in the $a_{l=0}^{I=0}$ partial wave at around 1.7 TeV \cite{Lee}. The $O(G_V^2)$ term helps to postpone unitarity violation to higher energies. This \emph{partial unitarity restoration} is most efficient for $G_V$ somewhat above the value
\beq
G_V^{\star}=v/\sqrt{3}\,,
\label{eq:GV}
\eeq
canceling the linear growth in $s$.\footnote{Recently, Ref.~\cite{Contino} discussed a more general criterion called ``partial UV completion" which determines (order of magnitude of) resonance couplings by saying that the resonance exchange should change the amplitude by $O(1)$ (rather than unitarize it). This approach is interesting to explore in the context of composite Higgs models, as in \cite{Contino}, where the amplitude grows more gently and there is no need to save it from violating unitarity right away. In Higgsless models the amplitude growth is more abrupt and there seems to be little room for a more general definition.}

Taking this mechanism at face value, unitarity of elastic $\pi\pi$ scattering can be preserved up to as high as 6(10) TeV for $M_V=2(1)$ TeV. Eventually, unitarity is violated by the logarithmically growing terms in $a_0^0(s)$. However, even before this, opening of inelastic channels like $\pi\pi\to VV$ has to be taken into account \cite{Papucci, Bar1,Falkowski}. Because of this, we will trust the amplitude \reef{eq:App} and the resulting partial unitarity restoration at most up to $\sim 2 M_V$.

Parameter $G_V$ also controls the width of the $V$ resonance:
\begin{equation}
\Gamma_V=\Gamma(V\rightarrow \pi\pi)=\frac{G_V^2M_V^3}{48\pi v^4}\,. 
\label{eq:GammaV}
\end{equation}
Interestingly, for $G_V\approx \bar G_V$ this width is an order of magnitude smaller than the Higgs boson width for the same mass, Eq.~\reef{eq:wH}. In this respect, vectors seem to be more efficient (albeit imperfect) unitarizers than scalars \cite{us}.

\begin{figure}[h]
\centering
\includegraphics[scale=0.4]{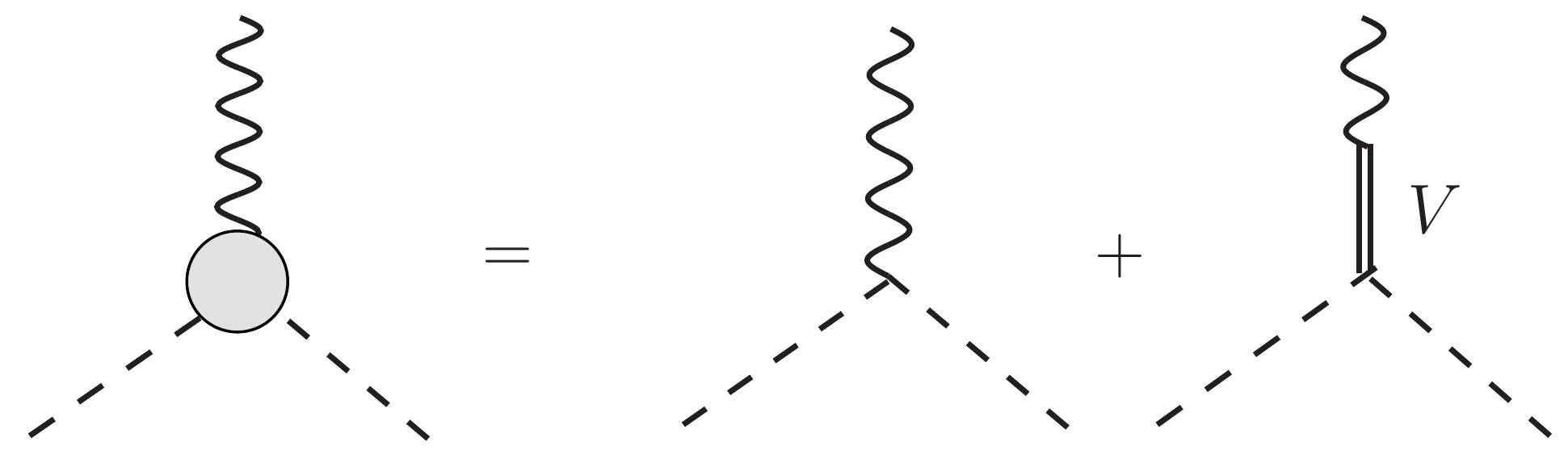}
\caption{Goldstone formfactor in Vector Meson Dominance.}
\label{formfactor}
\end{figure}

We next discuss parameter $F_V$, which measures the strength of $V$-gauge mixing. Among other things, this coupling controls the high $q^2$ behavior of the goldstone formfactor (see Fig.~\ref{formfactor}):
\begin{equation}
\langle\pi(p^{\prime})|J_V^{\mu}(q)|\pi(p)\rangle={\mathcal F}(q^{2})%
(p+p^{\prime})^{\mu}\,,\qquad {\mathcal F}(q^2)=1-\frac{F_VG_V}{v^2}\frac{q^2}{q^2-M_V^2}\,.
\label{eq:formfactor}
\end{equation}
If goldstones are composite objects, the formfactor is expected to go to zero at large $q^2$, which happens for
\beq
F_V G_V=v^2\,,\qquad {\mathcal F}(q^2)\to -\frac{M_V^2}{q^2-M_V^2}\,.
\label{eq:FVGV}
\eeq

Assuming that there is only one prominent resonance in the vector channel, relations \reef{eq:GV} and \reef{eq:FVGV} allow to determine both $G_V$ and $F_V$. These determinations, although approximate, are a step beyond NDA. We may be emboldened to take this step because, applied in QCD, this method gives the $\rho\pi\pi$ coupling and $\rho$-photon mixing in pretty good agreement with experiment \cite{Ecker0,Ecker,us}.\footnote{See also \cite{Pich} for a recent discussion of elastic unitarity in QCD pion scattering.} 

The role of couplings $F_{V,A}$ for the $S$ parameter will be discussed in the next section.

\section{$S$ parameter}
\label{sec:S}

Starting with \cite{Peskin}, most of the discussion of the electroweak precision tests in strong EWSB models has focused on the $S$ parameter. Here we review the necessary facts in some detail.

\subsection{Peskin-Takeuchi formula}
\label{sec:PT}

Consider the global symmetry current two point functions, which in the momentum space can be written as 
\begin{equation}
\int d^4x \ e^{-iqx}\left\langle 0\right|T(J^{a\mu}_V(x)J^{b\nu}_V(0))\left|0\right\rangle = -i\left(\eta^{\mu\nu}-\frac{q^\mu q^\nu}{q^2}\right)\Pi_{VV}(q^2)\delta^{ab}\,,
\end{equation}
and analogously for $AA$. Separating the goldstone pole in the axial channel, we write 
\begin{equation}
\begin{split}
\Pi_{AA}(q^2) &=v^2+q^2\widetilde{\Pi}_{AA}(q^2)\,, \\
\Pi_{VV}(q^2) &=q^2 \widetilde{\Pi}_{VV}(q^2)\,.
\end{split}
\end{equation}

Now, the $S$ parameter is related to the kinetic mixing of $W_3^\mu$ and $B^\mu$.
Since these couple to the currents $J^{3\mu}_{L,R}=\frac12(J^{3\mu}_{V}\pm J^{3\mu}_{A})$,
 it is tempting to identify
\begin{equation}
e_3\stackrel{\text{?}}{=}g^2\,\Pi'_{LR}(0)=\frac{g^2}{4}\left[\widetilde{\Pi}_{VV}(0)-\widetilde{\Pi}_{AA}(0)\right]\,. \label{eq:e3}
\end{equation} 
This equation cannot be exactly true since it treats gauge fields as nondynamical. Since the $W$ and $Z$ boson do propagate, their loops contribute to $e_3$, which is not captured by Eq.\ (\ref{eq:e3}).
However, this equation does capture the part of $e_3$ which comes from the presence of new degrees of freedom residing in the EWSB sector.
For a precise statement, consider the dispersion relation:
\begin{equation}
\widetilde{\Pi}_{VV}(q^2)=\int_0^\infty ds \ \frac{\rho_V(s)}{s-q^2+i\epsilon}, \ \ \rho_V(s)=\frac{1}{\pi}\text{Im}\, \widetilde{\Pi}_{VV}(s)\,, \label{eq:pivv}
\end{equation}
 and analogously for $AA$.
 Then Eq.\ (\ref{eq:e3}) can be rewritten as a spectral density integral:
 \begin{equation}
 e_3\stackrel{\text{?}}{=}\frac{g^2}{4} \int_0^\infty \frac{ds}{s}\left[\rho_V(s)-\rho_A(s)\right]\,.
 \end{equation}
 
 Whatever the EWSB model, there is one intermediate state which always contributes to this integral. This is the two goldstone state $\pi \pi$ in the vector channel 
 \begin{equation}
 \rho_V\supset \rho_{\pi\pi}(s)=\frac{1}{48\pi^2}\,.
 \end{equation}
 For the SM Higgs sector, this is in fact the full $\rho_V$ at one loop.
 Since $\rho_{\pi\pi}$ is constant in $s$, it gives a log-divergent contribution 
 \beq
 e_3(\pi\pi) = {g^2 \over 96\pi^2}\log {\Lambda \over \mu}\,.
 \label{eq:e3pipi}
 \eeq
 The IR part of the divergence will be cutoff at $\mu \sim M_Z$, since at these energies goldstones start mixing significantly with the $W$ and $Z$.
 The precise way in which this happens is not important for us. What is important however, is that this mixing and the resulting cancellation of the IR divergence is insensitive to high scales: it happens in precisely the same way in whatever heavy EWSB theory. Thus this effect will cancel in the $S$ parameter difference $e_3-e_3^{\rm SM}$ ($M_H\gg M_Z$), which frees us from having to compute it. 
 
 We finally arrive at the correct result:
 \begin{equation}
 \hat{S}=\frac{g^2}{4}\int_{\mu^2}^{\infty}\frac{ds}{s}\left[\left(\rho_V-\rho_A\right)-\left(\rho_V^{\rm SM}-\rho_A^{\rm SM}\right)\right]\,. \label{eq:Shat}
 \end{equation}
 The lower limit of integration must belong to the interval $M_Z^2\ll\mu^2\ll M_H^2$.
 It should also be much below whatever resonances of the strong EWSB model. Under these conditions, only $\pi\pi$ state matters for $s\sim\mu^2$. Its contribution will cancel among $\rho_V$ and $\rho_V^{\rm SM}$, and the integral will be independent of $\mu^2$. 
 
 
 The SM part of Eq.\ (\ref{eq:Shat}) can be fully evaluated. The UV divergence from $\rho_V=\rho_{\pi\pi}$ is canceled at $\Lambda \sim M_H$ by 
 \begin{equation}
\rho_A=\rho_{\pi h}=\frac{1}{48\pi^2}\left(1-\frac{M_H^2}{s}\right)^3\theta(s-M_H^2)\,.
\end{equation}
Evaluating also the finite term for completeness, we find an expression for the $S$ parameter in terms of the strong sector data:
\begin{equation}
 \hat{S}=\frac{g^2}{4}\int_{\mu^2}^{\infty}\frac{ds}{s}\left[\rho_V(s)-\rho_A(s)\right] -\frac{g^2}{96\pi^2}\left( \log \frac{M^{\text{ref}}_H}{\mu}+\frac{11}{12}\right)\,.
\end{equation}
Notice that the Higgs mass dependence is consistent with Eq.\ (\ref{eq:asympteps}).

This formula was originally derived and used in \cite{Peskin} to estimate the $S$ parameter in QCD-like Technicolor models, as a function of the numbers of technicolors and technifermions $N_\text{TC}$ and $N_\text{TF}$.  The point is that in QCD, the spectral densities $\rho_{V,A}$ can be extracted from experiment (measurements of hadroproduction in $e^+e^-$ collisions and $\tau$ decays). One can directly use these experimental densities to compute the $S$ parameter for $N_\text{TC}=3$ and $N_\text{TF}=2$, when Technicolor dynamics matches QCD. For other values of $N_\text{TC}$ and $N_\text{TF}$, Ref.~\cite{Peskin} argued that a reasonable estimate can still be obtained by rescaling the QCD densities appropriately.
 
\subsection{$S$ parameter and Vector Meson Dominance}
\label{sec:SVMD}

Our approach to computing $S$ will be different from \cite{Peskin} in two aspects. First, we will model spectral densities using the Vector Meson Dominance setup from section \ref{sec:VMD}, rather than taking the experimental QCD densities as a starting point. Second, we will allow for theories whose UV structure deviates significantly from QCD, as Walking or Conformal Technicolor, so that the approach of \cite{Peskin} is not applicable. In practice, this will mean that we will not enforce the second Weinberg sum rule; see section  \ref{sec:UVtail}.
 
Consider then the VMD Lagrangian  \reef{eq:LV}. Neglecting for the moment the $G_V$ coupling, the leading order spectral densities are given by
 \begin{gather}
\rho_V(s)=F_V^2 \delta(s-M_V^2)+\rho_{\pi\pi},\qquad \rho_A(s)=F_A^2 \delta(s-M_A^2)\,.
\label{eq:GV=0}
 \end{gather}
Here we are including the $\pi\pi$ state and the composite vectors in the zero width approximation. The resonances give a finite contribution\footnote{\label{note:4th}Notice that the vector and axial contributions are separately positive-definite as they should be according to the Peskin-Takeuchi formula. This is a nice feature of the antisymmetric tensor formalism. In the ``hidden local symmetry'' formalism there is an operator involving the vector resonance, $\text{Tr}(\rho_{\mu\nu} f^+_{\mu\nu})$ in the notation of Ref.~\cite{Contino}, which gives a non sign-definite contribution to $S$. In our opinion, this means that this operator always comes accompanied by a contact counterterm $\text{Tr}(f^+_{\mu\nu})^2$ so that the total $\Delta S>0$.}
\beq
e_3(\text{res})=\frac{g^2}{4}\left({F_V^2 \over M_V^2}-{F_A^2 \over M_A^2}\right)\,,
\eeq
to which we must add the UV divergent contribution \reef{eq:e3pipi} from $\rho_{\pi\pi}$. 

This log-divergence is however an artifact of neglecting $G_V$. The point is that the $\pi\pi$ spectral density is suppressed at $s\gg M_V$ by the destructive interference with the $V$ exchange diagram. This is the same mechanism which regulates the goldstone formfactor \reef{eq:formfactor}. Assuming the relation \reef{eq:FVGV}, the modified spectral density is
\beq
\tilde\rho_{\pi\pi}(s) = {1\over 48\pi^2}\left|{\mathcal F}(s)\right|^2= {1\over 48\pi^2}{M_V^4\over (s-M^2_V)^2+\Gamma_V^2M_V^2}.
\label{eq:pipimod}
\eeq
Notice that we took into account the finite width effect in the $V$ propagator, which was not necessary in the formfactor \reef{eq:formfactor} where $q^2$ is spacelike.  
Indeed, another consequence of having $G_V$ nonzero is that $V$ decays into $\pi\pi$, with a width given in Eq.~\reef{eq:GammaV}.

Since $V$ is no longer an asymptotic state, its pole contribution should not be added separately to the spectral density. Rather, this contribution is now described by the Breit-Wigner peak in $\tilde\rho_{\pi\pi}$. In particular, the integrated strength of this peak is exactly the same as of the delta function in \reef{eq:GV=0}. We conclude that Eq.~\reef{eq:pipimod} represents the full vector spectral density at $G_V\ne0$:
\beq
\rho_V(s)=\tilde\rho_{\pi\pi}(s)\qquad(G_V\ne0)\,.
\eeq
The dispersion integral is easy to evaluate in the approximation $\Gamma_V\ll M_V$, which remains reasonable for $M_V$ as high as 2 TeV. We find
\beq
e_3=\frac{g^2}{4}\left({F_V^2 \over M_V^2}-{F_A^2 \over M_A^2}\right)+{g^2 \over 96\pi^2}\left(\log {M_V \over \mu}+
O(1)\right)\,.
\eeq
Since $\tilde\rho_{\pi\pi}(s)$ goes to zero rapidly for $s\gg M_V$, it is at this scale that the UV logarithmic divergence is cut off. The finite correction to the logarithm could be easily evaluated, but we prefer not to show it explicitly. Unlike the general conclusion that the log-divergence is cutoff at $\Lambda\sim M_V$, this finite term would not be a robust prediction of this model. For example, it would change if we assumed an $s$-dependent width in the $V$ propagator.

Subtracting a reference SM contribution, we finally obtain the $S$ parameter:
\beq
\hat S=\frac{g^2}{4}\left({F_V^2 \over M_V^2}-{F_A^2 \over M_A^2}\right)+{g^2 \over 96\pi^2}\left(\log {M_V \over M^{\text{ref}}_H}+O(1)\right)\,.
\label{eq:SVMD}
\eeq
The logarithm can be set to zero by choosing the reference mass $M_H^{\text{ref}}=M_V$, and we will be left with the resonance contribution $\pm$ uncertainty of the order $g^2/96\pi^2\sim 0.5\times10^{-3}$.

We have based our discussion of the $S$ parameter on the dispersion relation. However, the goldstone contribution \reef{eq:e3pipi} is also simple to compute from the Feynman diagram
\begin{center}
\includegraphics[scale=0.75]{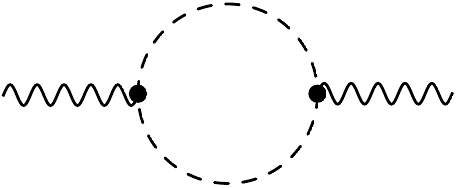}\raisebox{16pt}{\ .}
\end{center}
In the diagrammatic approach, adding $V$ exchanges amounts to including diagrams:
\begin{center}
\includegraphics[scale=0.75]{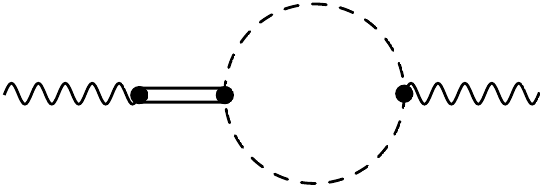}\raisebox{16pt}{\ ,}
\qquad
\includegraphics[scale=0.75]{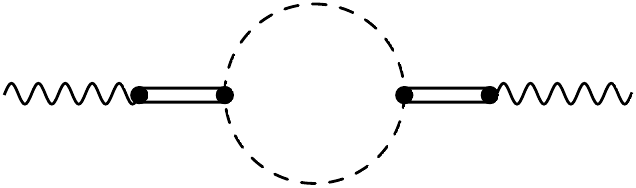} \raisebox{16pt}{\ .}
\end{center}
In this language, it would not be immediately clear how the $V$ exchanges help with canceling the $\log\Lambda$ dependence. In fact, it may seem that they make the situation worse because the added diagrams contain power-like divergences.  The resolution of the paradox lies in the fact that since these divergences are contained in the subdiagrams
\begin{center}
\includegraphics[scale=0.75]{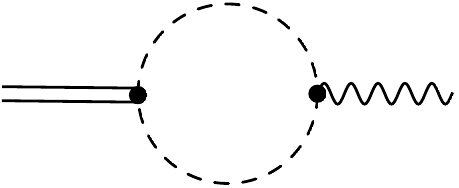}\raisebox{16pt}{\ ,}
\qquad
\includegraphics[scale=0.75]{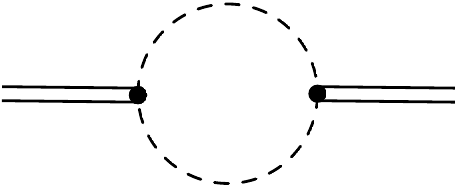}\raisebox{16pt}{\ ,}
\end{center}
they have to be interpreted as renormalization of $F_V$ and of the vector propagator.\footnote{\label{k0}Such a renormalization was discussed in \cite{Kamenik}. The residual quadratic divergences in Ref.~\cite{Kamenik} are related with contributions of the intermediate states other than $\pi\pi$, to be discussed below.}

Computing via the dispersion relation allows us not to deal with these issues. Renormalization is performed ``on the go", and we get a finite result in terms of renormalized parameters.

We will now comment on contributions of other intermediate states to the spectral densities. The $\pi V$ and $\pi A$ two particle states contribute to $\rho_A$ and $\rho_V$, respectively. In the large~$s$ limit these contributions are given by
\beq
\frac 1{24}\left (\frac{F_V-2G_V}{M_V}\right)^2\frac{s}{16\pi^2 v^2}\qquad \text{and}\qquad\frac 1{24}\left (\frac{F_A}{M_A}\right)^2\frac{s}{16\pi^2 v^2}\,,
\eeq
respectively, giving a quadratically divergent contribution to $S$. The origin of this is a bad UV behavior of the $\pi A$ and $\pi V$ formfactors, which will also cause a quadratic divergence in $T$ as we will see in section \ref{sec:TVMD} below. There, we will add two extra couplings $\kappa_A$ and $\kappa_V$ to the VMD Lagrangian, which will regulate the formfactors and will replace 
\beq
F_V-2G_V\to F_V-2G_V+2\kappa_V F_A,\qquad F_A\to F_A+2\kappa_A F_V
\eeq
in the spectral densities given above. In section \ref{sec:TVMD} we will fix the new couplings so that the quadratic divergence in $S$ (and simultaneously in $T$) vanishes. What remains is a logarithmically divergent $\Delta S$ which, we checked, is negligible in practice compared to the terms present in \reef{sec:SVMD}, if cut off at $\Lambda\sim 2 M_V$. (On the contrary, the contribution to $T$ will remain important even after cutting off the quadratic divergence.)

The $AV$ intermediate state appears in the spectral density $\rho_A$ as a result of the new couplings $\kappa_{A,V}$. For $s\gg M_V+M_A$ we have
\beq
\Delta \rho_A=\frac{s^2}{1536 \pi^2}\left(\frac{\kappa_A}{M_A^2}+\frac{\kappa_V}{M_V^2}\right)^2+O(s)\,,
\eeq
giving a \emph{quartically} divergent contribution in $S$.\footnote{In agreement with an observation made in \cite{Kamenik}.} However, this contribution remains reasonably small if cut off at $\Lambda\sim 2M_V$, due to phase space suppression. In a more realistic setup, the $AV$ formfactor would be softened giving an even smaller $\Delta S$.

Finally, the $VV$ and $AA$ intermediate states contribute to $\rho_V$. The relevant couplings are contained in the kinetic Lagrangian of the heavy spin-1 resonances \cite{Kamenik}.\footnote{\label{k1}We are grateful to J.~Kamenik and O.~Cata for pointing out our omission to discuss these intermediate states in the first version of the paper.} The corresponding spectral density is:
\beq
\Delta \rho_V(s) = \frac{1}{96 (4\pi)^2}\sum_{R=V,A} \left ( \frac s {M_R^2}+\frac 32\right)\left(1-\frac {4M_R^2}{s}\right)^{3/2}
\theta(s-4M_R^2)\,.
\eeq
Because of phase space, the contribution from $VV$ starts above $2M_V$, which is where our cutoff lies, so we do not include it.
The contribution from $AA$ should be included if $A$ is lighter than $V$, and it gives a quadratically divergent $\Delta S$:
\beq
\Delta S \sim \frac{g^2}{96\pi^2} \left(\frac{\Lambda}{8 M_A}\right)^2,\qquad \Lambda\sim 2 M_V.
\eeq
This is subdominant to the Goldstone loop contribution in \reef{eq:SVMD}, as long as the $M_V\lesssim 4 M_A$. The latter condition will be satisfied in most of the parameter space considered below in section \ref{sec:vs}.

To summarize, we find it reasonable to neglect contributions from other intermediate states to the $S$ parameter, and will keep only those giving rise to Eq.~\reef{sec:SVMD}.
\subsection{UV tail and Weinberg sum rules}
\label{sec:UVtail}

In this section, we collect several general remarks about the UV tail of the spectral densities entering the Peskin-Takeuchi formula. This can be inferred by considering the OPE of the $SU(2)_L$ and $SU(2)_R$ currents:
\beq
J^{a\mu}_{L}(x)J^{\dot{a}\nu}_{R}(0)\sim \left(\eta^{\mu\nu}\partial^2-\partial_{\mu}\partial_{\nu}\right){(x^2)^{\Delta_\Phi/2-2}}\,\Phi^{a\dot{a}}(0)\,.
\label{eq:OPE}
\eeq 
The scalar $\Phi^{a\dot{a}}$ is the lowest dimension scalar transforming in the bi-adjoint of $SU(2)_L\times SU(2)_R$.
As we will see below, the dimension $\Delta_\Phi$ will control the UV behavior of the dispersive integral.

If the UV theory is weakly coupled, operator $\Phi$ and its dimension can be found explicitly. For example:
\beq
\begin{split} 
\text{Technicolor:}\quad\Phi &= \left(\bar{\psi}P_LT^a\psi\right)\left(\bar{\psi}P_RT^{\dot a}\psi\right)\qquad (\Delta_\Phi=6)\,, \\
\text{SM:}\quad\Phi &= \text{tr} \left[\mathbb {H}T^a_L\mathbb{H}^{\dagger}T^{\dot{a}}_R\right]\qquad (\Delta_\Phi=2), 
\end{split} \label{eq:Phi's}
\eeq
where $\mathbb H$ is the SM Higgs field in the matrix bi-doublet notation. For a strongly coupled UV theory, like Walking \cite{Holdom} or Conformal \cite{Luty} Technicolor, the dimension of $\Phi$  is in general unknown, except for the lower bound $\Delta_\Phi>1$ as a consequence of unitarity.  
 Interestingly, under the special assumption made in Conformal Technicolor (scalar bi-doublet of dimension close to 1), one can also derive an \textit{upper} bound 
 \beq
 \Delta_\Phi\lesssim 2+2.85(\Delta_H-1)<4\qquad\text{(Conformal Technicolor)}\,. 
 \label{eq:CTCbound}
 \eeq
 See \cite{V,DDV} for the actual bound, and \cite{R} for the theory behind.
 
Coming back to the OPE \reef{eq:OPE}, it determines the UV asymptotics of the LR self-energy:
 \beq
 \Pi_{LR}(q^2)\sim {(-q^2)^{1-\Delta_\Phi/2}}{\left\langle \Phi\right\rangle},\label{eq:OPE2}
 \eeq
times a factor of $\log(-q^2)$ if $\Delta_\Phi$ is an even integer. The VEV of $\Phi$ is nonzero since conformal symmetry, and the global symmetry, are broken in the IR; by dimensional analysis:\footnote{We are not trying to keep track of factors of $\pi$.}
 \beq
 \left\langle \Phi\right\rangle \sim (\Lambda_{EW})^{\Delta_\Phi}\,.
\eeq
The sign of $\left\langle \Phi\right\rangle$ is undetermined in general, except for Technicolor/Walking Technicolor theories based on a vector-like gauge theory in the UV. For such theories Witten \cite{W} has shown that $\Pi_{LR}>0$ in the Euclidean $q^2\rightarrow-q_E^2$, $q_E^2>0$. Thus necessarily $\left\langle \Phi\right\rangle>0$ in this case \cite{Sundrum}.

From Eq.~(\ref{eq:OPE2}), we conclude:
\beq
\rho_V(s)-\rho_A(s)\sim s^{-\Delta_\Phi/2}\qquad(s \gg \Lambda_{EW}^2)\,.
\eeq
This implies that the Peskin-Takeuchi formula always converges in the UV. Another way to see this absence of the UV sensitivity is to notice that the operator 
\beq
\mathcal{O}=W_{\mu\nu}^aB_{\mu\nu}\,\Phi^{a3} ,
\eeq
interpolating the $S$ parameter effective operator $\langle U\hat{W}_{\mu\nu}U^{\dagger}\hat{B}_{\mu\nu}\rangle$, is always irrelevant since $\Delta_\Phi>1$.

Finally, we discuss the first and second Weinberg sum rules \cite{Weinberg}
\beq
\begin{split}
& \int_0^\infty ds \left[ \rho_V(s)-\rho_A(s)\right]=v^2\qquad(\Delta_\Phi > 2)\,,\\
& \int_0^\infty ds\, s\left[ \rho_V(s)-\rho_A(s)\right]=0\qquad(\Delta_\Phi >4)\,,
\end{split}
\eeq
which follow by expanding the dispersion relations \reef{eq:pivv} at $q^2\rightarrow \infty$ and setting the coefficients to zero. This is legitimate as long as $\Pi_{LR}(q^2)$ decays fast enough, which happens for $\Delta_\Phi$ above indicated values.

For example, none of the two sum rules hold for the SM ($\Delta_\Phi=2$) while both of them would be valid for Technicolor ($\Delta_\Phi=6$).

Most of the Walking Technicolor literature assumes that the fermion bilinear dimension gets close to 2 in the walking regime. Then if operator dimensions are assumed to factorize (unjustified approximation), the four-fermion operator $\Phi$ as in \reef{eq:Phi's} will have dimension close to 4. The second Weinberg sum rule then gets an important log-divergent contribution from the walking regime, cut off at the scale where the theory transitions to the weakly coupled UV gauge theory. One can try to model this transition and estimate the resulting contribution \cite{Appel}.

For Conformal Technicolor models, we can rely on the rigorous bound \reef{eq:CTCbound} which implies that $\Delta_\Phi  < 4$ in the interesting range of $\Delta_H$. This means that the second Weinberg sum rule will have a \emph{powerlike} divergence, and cannot be used.\footnote{If the model transitions to a weakly coupled gauge theory in the deep UV, the divergent contribution will be cut off at the transition scale and cancelled by a threshold correction of the opposite sign, leaving a finite remainder which seems impossible to predict with current technology.}

Applied to the VMD spectral densities \reef{eq:GV=0}, the Weinberg sum rules would imply
\beq
\begin{split}
&F_V^2-F_A^2=v^2+O\left({M_V^2}/{48\pi^2}\right)\,, \label{eq:W1}\\
&F_V^2M_V^2-F_A^2M_A^2=O\left({M_V^4}/{48\pi^2}\right)\,, 
\end{split} 
\eeq
where $O\left(...\right) $ represent the contribution from $\pi\pi$, negligible for $M_V\lesssim 2$ TeV.
These equations have to be taken with a grain of salt. They can be easily disturbed by presence of multiple resonances or, as we have seen, the original sum rule could be simply invalid. This is especially true of the second equation, strongly tilted into the heavy part of the spectrum.

When comparing with the data in section \ref{sec:vs}, we will require that the VMD spectrum satisfy the first Weinberg sum rule. The second sum rule will not be imposed; it will turn out to hold only in a small region of the full parameter space, which will be different from the region preferred by the EWPT.

\section{$T$ parameter}
\label{sec:T}

In the SM, the $T$ parameter at one loop gets important contributions $O(y_t^2)$ and $O(g'^2)$, since these are the two largest couplings breaking the custodial symmetry. 
 In strong EWSB, the $O(y_t^2)$ contribution may get modified because of top quark compositeness or other effects \cite{SILH,Luty2}, and there is an extensive literature trying to keep track of these modifications in explicit models \cite{Santiago,extended, extended1,Pomarol,Duccio,Chivukula-top}.
 
 On the other hand, the $O(g'^2)$ contribution is typically included in the analysis by just keeping the chiral logarithm. There is only a handful of papers which discuss the role of resonances for the $T$ parameter \cite{Chivukula-long,Chivukula,Abe:2008hb,us,Kamenik}. This can be contrasted to the $S$ parameter, where it is standard to keep both the chiral log and the contribution of resonances. In this paper, we will follow \cite{us} and try to investigate the resonance contribution to $T$.
 
 \subsection{$T$ parameter and goldstone wavefunction renormalization}
\label{sec:gold}

 At leading order in $g'$, the $T$ parameter can be determined from the EWSB sector data by means of the equation:
\beq
\left.\left\langle J^3_L(q)J^3_L(-q)\right\rangle-\left\langle J^+_L(q)J^-_L(-q)\right\rangle \right|_{q\rightarrow 0} = -i \left(\eta^{\mu\nu}-\frac{q^\mu q^\nu}{q^2}\right)\frac{v^2}{4}e_1\,. \label{eq:e1def}
\eeq
Since the SM gauge fields couple to the EWSB sector by\footnote{Here \ldots\ stands for terms quadratic in gauge fields, needed to have full gauge invariance to $O\left(g^2,g'^2\right)$.}
\beq
\Delta\mathcal{L}=gW^a_\mu J^{a\mu}_L+g'B_\mu J^{3\mu}_R+\ldots,\nonumber
\eeq
it is clear that Eq.\ \reef{eq:e1def} is consistent with the definition of $e_1$ in section \ref{sec:brief}.

 Since $g$ does not break the custodial symmetry, we can set it to zero and view $SU(2)_W$ gauge fields as non-dynamical. However, $B_\mu$ does propagate, and this makes the study of $T$ more difficult than that of $S$, where all gauge dynamics could be switched off.
 
One complication in interpreting and applying Eq.\ \reef{eq:e1def} is caused by the fact that the propagating $B_\mu$ mixes with $\pi^3$ (except in the Landau gauge; see below), shifting the massless pole in the $\left\langle J^3_L J^3_L\right\rangle$ correlator. One simple and effective way to prevent this from happening is to impose an IR cutoff which turns off $B^\mu$ modes below some energy $\mu$. This eliminates $B^\mu$-$\pi^3$ mixing at zero momentum and, simultaneously, regulates the IR logarithmic divergence present in $e_1$ (see below). The introduced $\mu$ dependence will cancel when subtracting $e_1^{\rm SM}$, just as it happened for the $S$ parameter.

 The L/R current expansion starts with 
\beq
J_{L,R}^a=\frac{v}{2}\partial_\mu\pi^a\pm\frac{1}{2} \epsilon^{abc}\pi^b\partial_\mu \pi^c + \ldots ,
\eeq
where \ldots\ contains higher-order terms as well as terms involving other fields. E.g.~in the SM
\beq
J_{L,R}^a \supset \pm \frac{1}{2}h\,\partial_\mu \pi^a\qquad\text{(SM)}\,.
\label{eq:JLSM}
\eeq
At $O(g'^2)$, goldstone contributions to the current-current correlators in Eq.\ \reef{eq:e1def} are given by three types of diagrams ($\otimes= J_L$, dashed lines = goldstones, wavy lines = $B_\mu$): 
\begin{equation}
\includegraphics[scale=0.3]{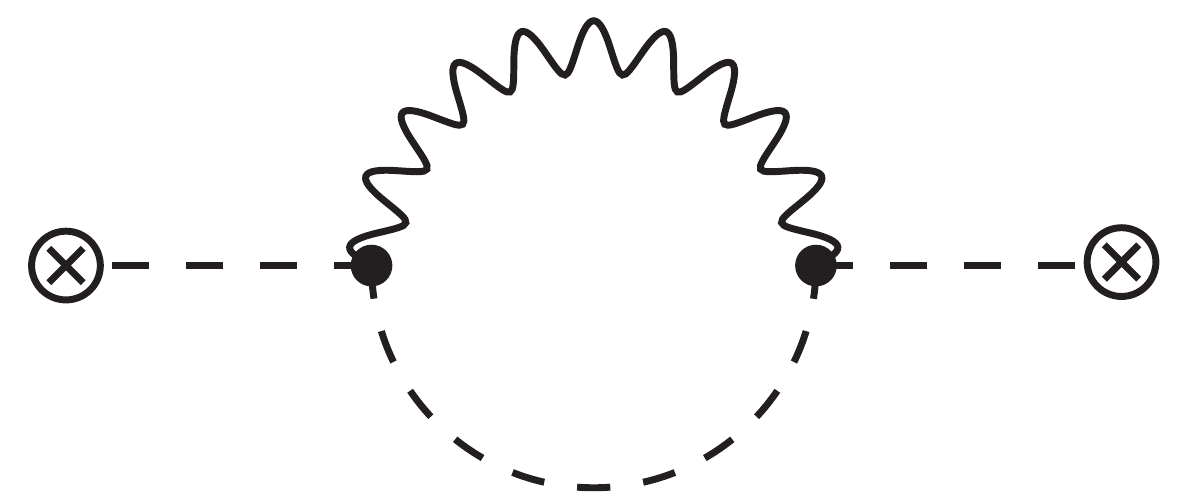}
\qquad\qquad
\includegraphics[scale=0.3]{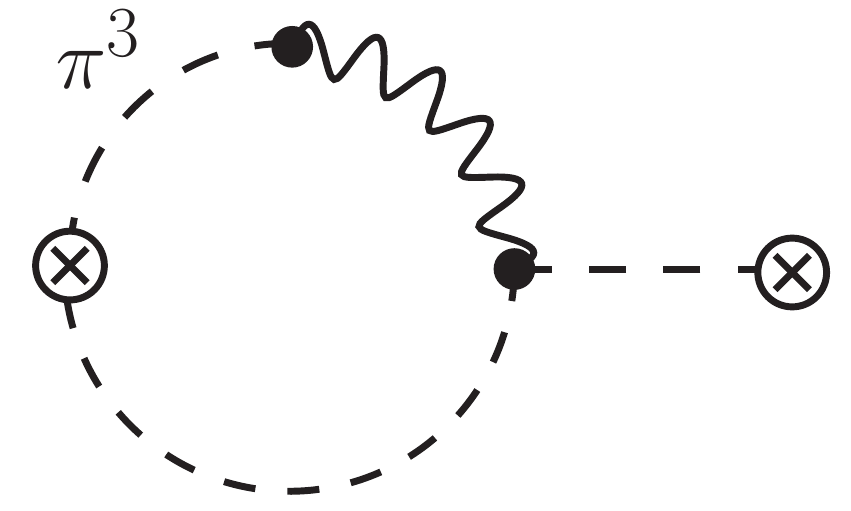}
\qquad\qquad
\includegraphics[scale=0.3]{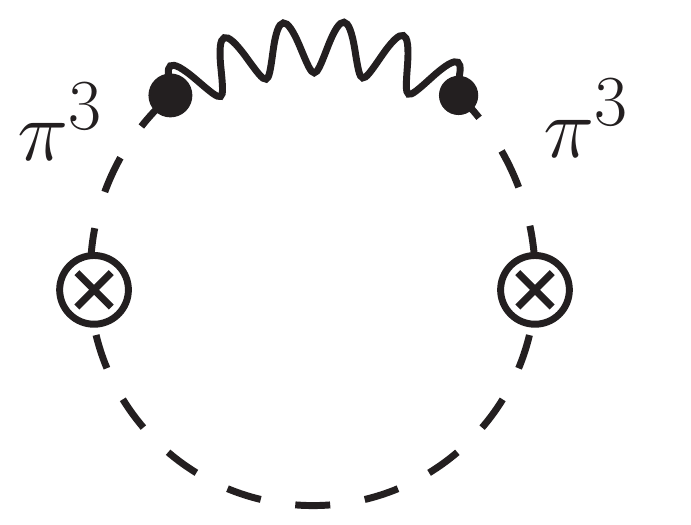} 
\label{eq:JLJL}
\end{equation}

In the Landau gauge, the kinetic $B^\mu$-$\pi^3$ mixing vanishes, and only diagrams of the first type survive. In the $q\rightarrow 0$ limit, these diagrams can be interpreted as corrections to the goldstone wavefunction renormalizations $Z$.
We thus arrive at a very useful relation \cite{Longhitano,Georgi,Barbieri-Cargese,Barbieri-2loop,Barbieri-book}:
\beq
e_1=\delta Z_+-\delta Z_3|_{\text{Landau gauge}}\qquad\text{(one loop)}\,.
\label{eq:Longhitano}
\eeq
Notice however that, in principle, we could compute $e_1$ also in other gauges, as long as we include all diagrams in Eq.\ \reef{eq:JLJL}.\footnote{Alternatively one can renormalize the chiral Lagrangian while working in the background field gauge, which is necessary to preserve the chiral symmetry \cite{Chivukula}.}

In a generic heavy model, we have
\beq
\delta Z_+ = -\frac{3 g'^2}{32 \pi^2}\log \frac{\Lambda}{\mu}\,,
\eeq
while in the SM we also have, from the diagram with the Higgs boson and $B_\mu$ circulating in the loop,
\beq
\delta Z_3 = -\frac{3 g'^2}{32 \pi^2}\log \frac{\Lambda}{M_H}\qquad\text{(SM)}\,.
\eeq
From here we can verify the $e_1^{\rm SM}$ behavior given in \reef{eq:asympteps} and obtain the well-known goldstone contribution to the $T$ parameter in a heavy model without Higgs:
\beq
\hat T = -\frac{3 g'^2}{32 \pi^2}\log \frac{\Lambda}{M^{\text{ref}}_H}\qquad\text{(Higgsless; from goldstones)}\,.
\label{eq:Tgold}
\eeq
 
 \subsection{Resonance contribution and UV sensitivity}
 \label{sec:TUV}
 
Eq.~\reef{eq:Tgold} brings out several questions. At what scale is the logarithmic divergence cut off? Can it be cut off by the resonances, as it happened for the $S$ parameter? Can we estimate the finite contribution from the resonances, as we did for $S$? 

In the low energy effective theory, the resonance contribution to the $T$ parameter can be estimated by adding to the chiral Lagrangian an effective operator
 \beq
 \frac{g'^2}{16 \pi^2v^2} \langle U (D_\mu U)^\dagger T^3\rangle^2\,,
 \label{eq:TNDAop}
 \eeq
 where we indicated the coefficient of the size expected from NDA. The logic is that such an operator can be generated by integrating out the resonances. The resulting order of magnitude estimate for $T$ is  
 \beq
T_{\text{res}} = O(g'^2/16 \pi^2)\,,
 \label{eq:TNDA}
 \eeq
which looks formally subleading to the logarithmically enhanced goldstone contribution \reef{eq:Tgold}. Still, it is interesting to ask whether the resonance contribution may become numerically large; this will be discussed in the next section in the context of VMD.
 
The above discussion presupposes that the $T$ parameter is not sensitive to the deep UV scales, where the EWSB sector is assumed to sit at a conformal fixed point. This, however, is not automatic, even if the UV fixed point is custodially-symmetric. We have to impose an additional assumption that the UV theory should not contain any relevant scalar operator $\Xi$ which is singlet under $SU(2)_W\times U(1)_Y$ but not under the full $SU(2)_L\times SU(2)_R$ group. For example, the $T_L=0$, $T_R^3=0$ component of a $(\mathbf{1},\mathbf{n})\oplus (\mathbf{n},\mathbf{1})$ representation of $SU(2)_L\times SU(2)_R$ could be such a scalar ($\mathbf{n}$ must be odd to have a $T_R^3=0$ component). 

If a relevant $\Xi$ were present in the theory, it could be generated when the EWSB theory is coupled to the SM. The coefficient of this operator would then grow in the IR. If $\Xi$ were strongly relevant, this would bring back the hierarchy problem. If $\Xi$ were weakly relevant, then a large hierarchy could in principle be obtained. However, the IR theory in this case would be custodial-symmetry violating, predicting an unacceptably large value of the $T$ parameter.

In weakly coupled UV theories, the lowest dimension operators with the $\Xi$ quantum numbers are ($\mathbf{n}=\mathbf{3}$ in both cases):
\beq
\begin{split} 
\text{Technicolor:}\quad\Xi &= \left(\bar{\psi}P_L\psi\right)\left(\bar{\psi}P_RT^{\dot a}\psi\right)\qquad (\Delta_\Xi=6)\,, \\
\text{SM:}\quad\Xi &= |H^\dagger D_\mu H|^2 \qquad (\Delta_\Xi=6)\,, 
\end{split} \nonumber
\eeq
and are irrelevant, so that the $T$ parameter is not UV sensitive. But in a general theory this may not be true. The requirement that no relevant $\Xi$ be present should be added to the list of conditions on a viable Conformal Technicolor model. At present it is not known if this is a serious constraint. The existing studies of Conformal Technicolor viability \cite{R,V,DDV} were based on the OPE of a $(\mathbf{2},\mathbf{2})$ with itself, which does not contain a scalar with the $\Xi$ quantum numbers. 

 \subsection{$T$ parameter and Vector Meson Dominance}
 \label{sec:TVMD}
 
 We now come to the question of computing the resonance contribution to the $T$ parameter. One may wonder if such a computation could be done starting from a dispersion representation, as was done for $S$. Eq.~\reef{eq:e1def} shows that the $T$ parameter is naturally related to the \emph{four point} function of two left and two right currents, which indicates that finding such a representation is more complicated than for $S$, where the relevant object was a two point function. Interestingly, there exists a simple dispersion relation for the electromagnetic pion mass difference in QCD \cite{Das}. In particular, in that case current four point functions can be reduced to two point functions using PCAC and current algebra. Since the $T$ parameter is related to the pion wavefunction difference via Eq.~\reef{eq:Longhitano}, it is tempting to think that a simple dispersion relation may exist here as well.

 Still, at present we do not know any such dispersion relation for the $T$ parameter.\footnote{See however \cite{Nutbrown:1970im} for old work on dispersion relations for current four point functions.} For this reason we will evaluate $T$ by using a diagrammatic approach, starting from the VMD Lagrangian introduced in section \ref{sec:VMD}. Eq.~\reef{eq:Longhitano} says that we must compute goldstone wavefunction renormalizations due to $B_\mu$ exchange in the Landau gauge. Only $\delta Z_+$ is nonzero; it is given by the two diagrams
 \begin{center}
\includegraphics[scale=0.4]{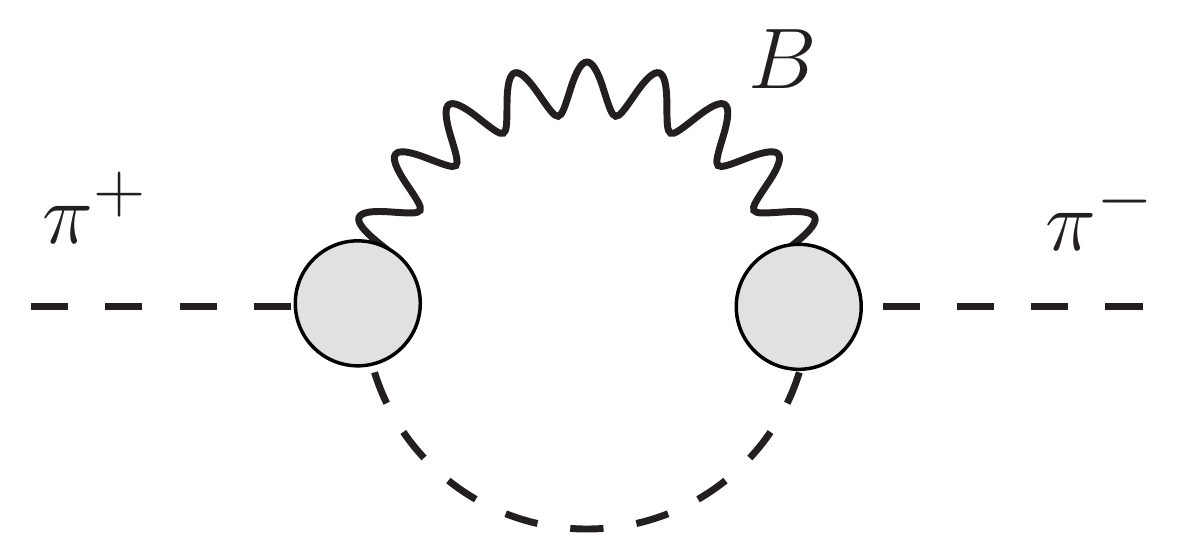}(a)
\qquad\raisebox{2em}{$+$}\qquad 
\includegraphics[scale=0.4]{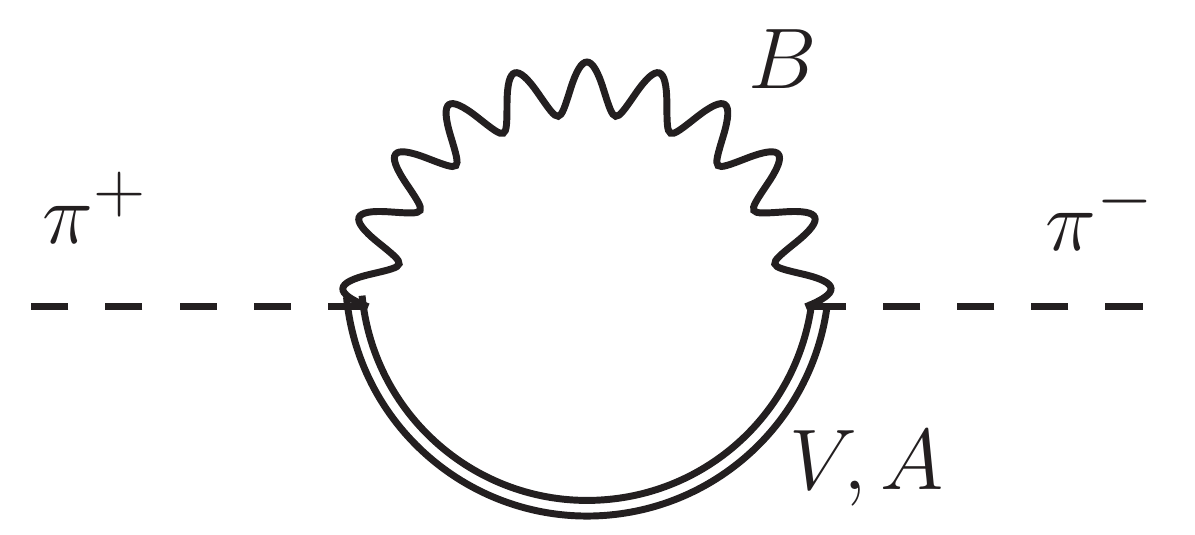}(b)
\end{center}

The blobs in diagram (a) stand for the $\pi\pi B_\mu$ vertex softened by the $V$ exchange diagram, as in Fig.~\ref{formfactor}. 
This vertex is an off-shell continuation of the pion formfactor but is actually given by the same formula, modulo $O(q^\mu)$ terms not contributing in the Landau gauge:
\vspace{-0.5em}
\begin{center}
\includegraphics[scale=0.4]{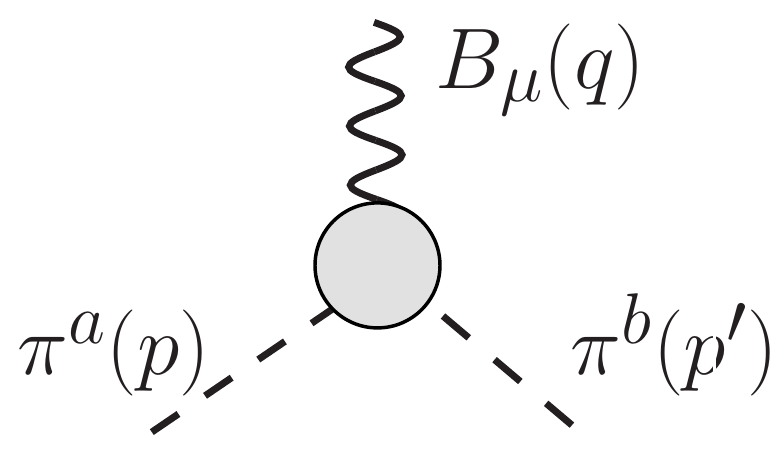}\qquad\raisebox{2em}{$\displaystyle=\frac{g'}{2}\epsilon^{ab3}\left(1-\frac{F_V G_V}{v^2}\frac{q^2}{q^2-M_V^2}\right)\left(p+p'\right)^\mu+O(q^\mu)$\,.}
\end{center}
Diagram (a) contribution to $\delta Z_+$ is then given by
\beq
\delta Z_+\text{(a)}=-\frac{3g'^2}{32\pi^2}\left[\left(1-\frac{F_V G_V}{v^2}\right)^2 \log\frac{\Lambda}{M_V}+\log\frac{M_V}{\mu}-\frac{F_V^2 G_V^2}{2v^4} \right]+O(M_V^2/\Lambda^2)\,.
\label{eq:3a}
\eeq
where we regulate the IR divergence as in section \ref{sec:gold}. This formula is easy to understand: the structure of IR and UV log-divergent terms is dictated by the low and high energy limits of the pion formfactor. Under condition \reef{eq:FVGV}, the formfactor asymptotes to zero and the UV divergence disappears.

On the other hand, diagram (b) gives a contribution which is, generically, quadratically divergent:\footnote{\label{k15}This is factor 2 bigger than the quadratic divergence reported in \cite{us} and \cite{Kamenik}. This should be due to an inconsistent normalization of the heavy vector propagator used in those works.}
\beq
\delta Z_+\text{(b)}=\frac{3g'^2}{8}\left[\frac{\left(F_V-2G_V\right)^2}{M_V^2}+\frac{F_A^2}{M_A^2}\right]\frac{\Lambda^2}{16\pi v^2}+\ldots
\label{eq:Tquad}
\eeq
This quadratic divergence is due to the $O(k^0)$ part of the heavy vector propagator. This most singular part happens to be transverse, as can be seen by rewriting the propagator as:
\begin{gather}
\Delta_{\mu\nu,\rho\sigma}=-2i\frac{g_{\mu\rho}g_{\nu\sigma}-g_{\nu\rho}%
g_{\mu\sigma}+M_R^{-2} P_{\mu\nu,\rho\sigma}}{k^{2}-M_R^{2}}\,,\label{eq:Rprop1}
\\
P_{\mu\nu,\rho\sigma}=-g_{\mu\rho}g_{\nu\sigma}k^2+g_{\mu\rho} k_\nu k_\sigma-g_{\mu\sigma} k_\nu k_\rho-(\mu\leftrightarrow\nu),\qquad k^\mu P_{\mu\nu,\rho\sigma}=0\,.\nn
\end{gather}
This in turn means that, for extracting quadratic divergence, it is sufficient to compute the $\pi V B$ and $\pi A B$ vertices modulo $ O(\del_\mu V^{\mu\nu}, \del_\mu A^{\mu\nu})$. Expanding \reef{eq:LV} and integrating by parts when necessary, the corresponding cubic couplings are given by:
\beq
\frac{F_V-2G_V}{\sqrt{2} v} \langle V^{\mu\nu} [\del_\mu \hat\pi,\hat B_\nu]\rangle+\frac{F_A}{\sqrt{2} v} \langle A^{\mu\nu} [\del_\mu \hat\pi,\hat B_\nu]\rangle + O(\del_\mu V^{\mu\nu}, \del_\mu A^{\mu\nu})\,.
\label{eq:piAV}
\eeq
This explains the coefficients in \reef{eq:Tquad}.

One can take two points of view with respect to the quadratic divergence. The first one would be to consider only those values of parameters for which the $\Lambda^2$ term vanishes. In this case, diagram (b) contributes with a negative logarithmic term:
\beq
\delta Z_+\text{(b)}|_{F_V=2G_V,F_A=0}=
-\frac{3g'^2}{32\pi^2}\frac{G_V^2}{v^2}\log \frac{\Lambda}{M_V} +O(1)
\label{eq:3blog}
\eeq
This situation is realized e.g.~in the well-known three-site model which has $G_V=v/2$, $F_V=v$. The $T$-parameter in the three-site model has been studied in \cite{Chivukula-long,Chivukula}; the sum of Eq.~\reef{eq:3a} and \reef{eq:3blog} agrees with their result.\footnote{\label{k2}An extra log-divergent term claimed in \cite{Kamenik} is in fact an error. We thank J.~Kamenik for discussions.} 

The second point of view would be to accept that in general $F_V\ne 2 G_V$, $F_A\ne 0$. The quadratic divergence then is a potentially physical effect, providing a positive contribution to $T$, which may perhaps improve the electroweak fit \cite{us}. However, to assess its importance it is interesting to know a mechanism by which it may be cut off. As one possible mechanism, Ref.~\cite{us} has proposed to add to the Lagrangian two extra parity-invariant terms:
\beq
i\kappa_{A}\langle A^{\mu\nu}[\nabla_{\rho}V^{\rho
\nu},u_{\mu}]\rangle+i\kappa_{V}\langle V^{\mu\nu}[\nabla_{\rho}A^{\rho\nu
},u_{\mu}]\rangle\,. \label{eq:L2V}%
\eeq
The effect of these terms on the $T$ parameter is to replace diagram (b) by
\begin{center}
\includegraphics[scale=0.4]{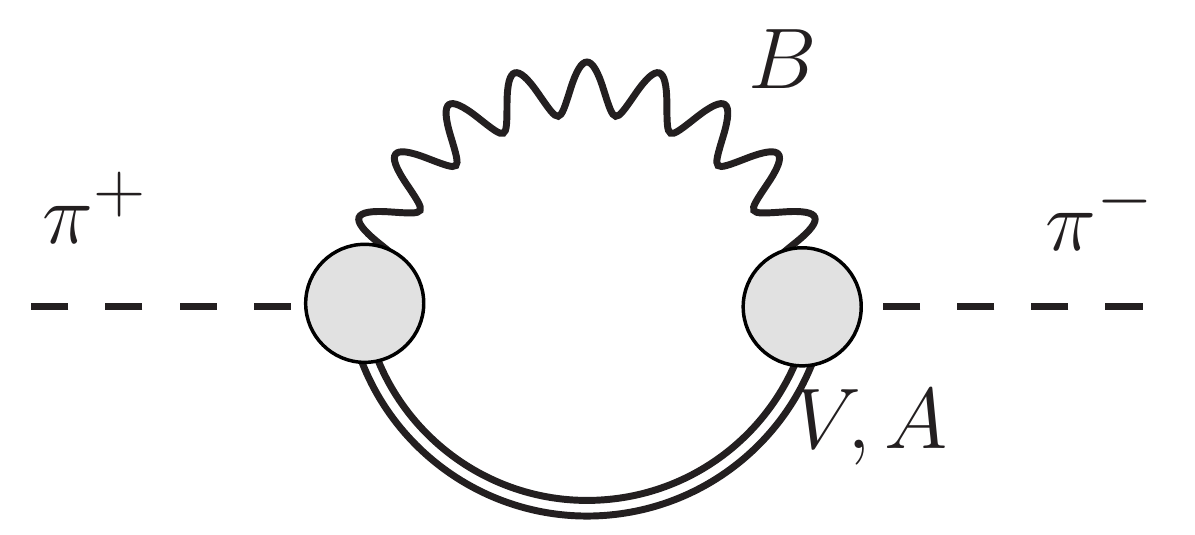}(b$'$)
\end{center}
Here the blobs stand for formfactors: the sum of the direct coupling present already in diagram (b), and the coupling where $B_\mu$ first mixes with $V(A)$ which then couples to $\pi A$ ($\pi V$) via \reef{eq:L2V}.

The formula for the quadratic divergence is now modified to
\beq
\delta Z_+\text{(b$'$)}=\frac{3g'^2}{8}\left[\frac{\left(F_V-2G_V+2\kappa_VF_A\right)^2}{M_V^2}+\frac{(F_A+2\kappa_AF_V)^2}{M_A^2}\right]\frac{\Lambda^2}{16\pi v^2}+\ldots
\label{eq:b'quad}
\eeq
This can be understood as follows. What matters for the quadratic divergence is the formfactor behavior at large $B_\mu$ momenta. In this case, the vector meson with which $B_\mu$ mixes can be integrated out. This gives cubic couplings as in \reef{eq:piAV} with $F_V-2G_V$ and $F_A$ shifted as in \reef{eq:b'quad}.

In what follows we will assume that the couplings $\kappa_{A,V}$ are such that the formfactors approach zero at large $B_\mu$ momenta:\footnote{There is a sign discrepancy with App.\ A of \cite{us}.}
\beq
F_V-2G_V+2\kappa_VF_A=0,\qquad F_A+2\kappa_AF_V=0\,,\label{eq:kappaVA}
\eeq
making the $\Lambda^2$ term vanish. 

With the quadratic divergence cancelled, one can go ahead and compute the logarithmic and finite terms. The formula is given in Appendix \ref{sec:Tbulky} and the results for comparison with the electroweak precision data will be reported in the next section.

We would now like to include other cubic $\pi AV$ couplings. Up to integration by parts, there are four other terms that one can write down, differing from \reef{eq:L2V} by the order in which indices are contracted: 
\begin{gather}
\mathscr{O}_1=\left\langle \nabla_{\rho}A^{\mu\nu}\left[V^{\mu\nu},u^{\rho}\right]\right\rangle\,,\quad  \mathscr{O}_2=\left\langle V^{\mu\nu}\left[A^{\mu\nu},\nabla_{\rho}u^{\rho}\right]\right\rangle\,,\nn\\
\mathscr{O}_3=\left\langle A^{\mu\nu}[V^{\rho\nu},\nabla_{\rho} u_{\mu}]\right\rangle\,,
\quad  \mathscr{O}_4=\left\langle V^{\mu\nu}[A^{\rho\nu},\nabla_{\rho} u_{\mu}]\right\rangle\,. 
\label{eq:newops}
\end{gather}
The effect of these couplings on the $T$ parameter is discussed in detail in Appendix \ref{sec:piAV}. For various reasons only the first of these operators turns out to give a nonzero contribution to $T$. This contribution is only logarithmically divergent; its numerical importance will also be studied in the next section.

\section{Theory vs Data}
\label{sec:vs}

We will now make a numerical comparison of the Vector Meson Dominance model with the electroweak precision data.

\begin{figure}[htbp]
\begin{center}
\includegraphics[scale=0.7]{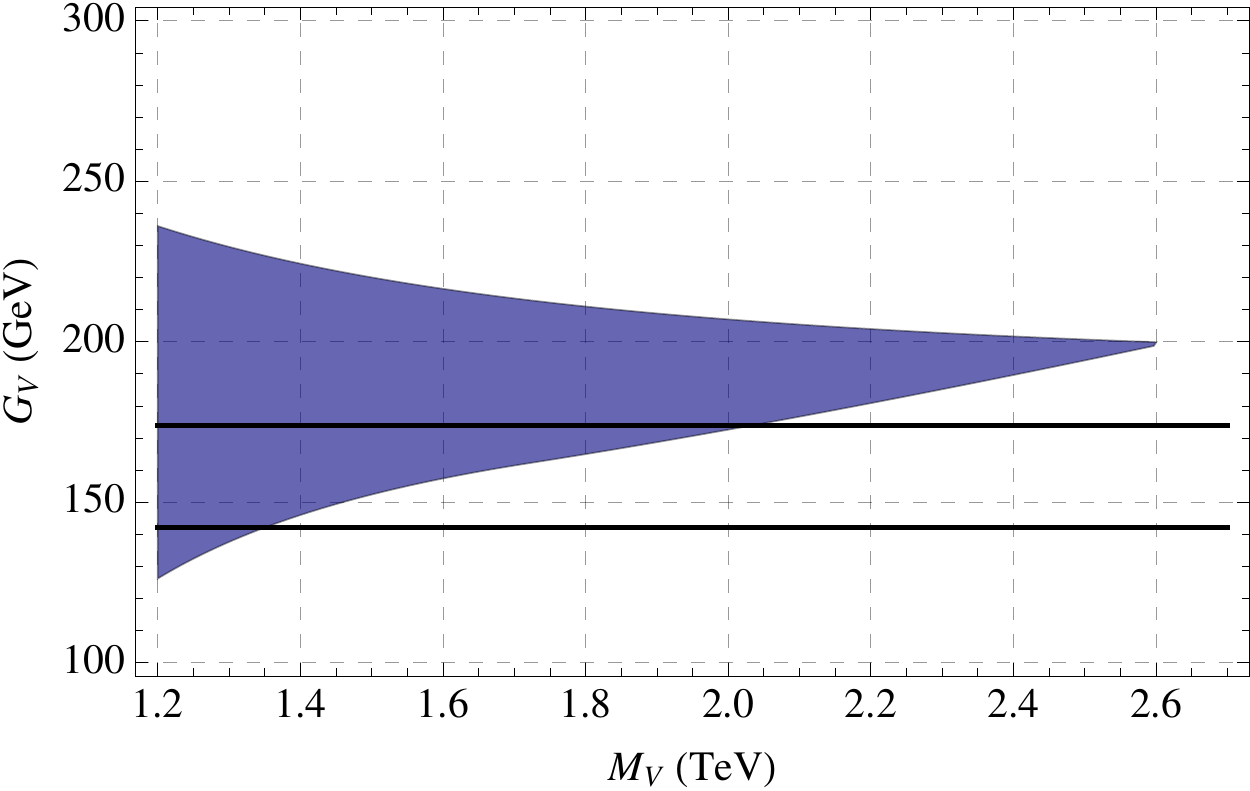}
\caption{The shaded region gives a constraint on $G_V$ from imposing elastic unitarity in the $\pi\pi$ scattering amplitude up to the energy of $2M_V$. 
The $v/\sqrt{2}$ and $v/\sqrt{3}$ lines are given for comparison.}
\label{fig:GVMV}
\end{center}
\end{figure}

The first step is to impose the constraint of elastic unitarity in $\pi\pi$ scattering, discussed in section \ref{sec:VMD}. We choose to impose that the $a^0_0$ partial wave remain less than 1 up to the energy $2M_V$. Other partial waves give weaker constraints. For each $M_V$, we obtain a range of allowed $G_V$, see Fig.~\ref{fig:GVMV}. We vary $M_V$ from 1.2 to 2.6 TeV. For larger $M_V$ it becomes impossible to maintain unitarity for whatever value of $G_V$, while smaller values of $M_V$ would give too large $S$ parameter (see below). We see that the $G_V$ preferred by this argument are indeed somewhat bigger than $v/\sqrt{3}$ from \reef{eq:GV}. The central value is not far from $G_V=v/\sqrt{2}$, which would be chosen if we imposed the pion formfactor constraint \reef{eq:FVGV} together with the $F_V=2G_V$ relation.

We note that Ref.~\cite{us} used a fixed 3 TeV cutoff. We find an $M_V$ dependent cutoff, like the one used here, more reasonable, since the resonance $V$ better be able to cure the ills of the theory without further help in a significant range of energies, for the logic of Vector Meson Dominance to make sense. 

For each $G_V$ from the allowed range, we will fix $F_V$ via Eq. \reef{eq:FVGV} to have a well-behaved pion formfactor, Eq.~\reef{eq:FVGV}. We then fix $F_A$ using the first Weinberg sum rule \reef{eq:W1}, neglecting the $\pi\pi$ contribution in the RHS. For reasons explained in section \ref{sec:UVtail}, we do not impose the second Weinberg sum rule. If we were to impose it, the resulting value of $M_A$ would give a large positive $S$ parameter, excluding the model.

\begin{figure}[htbp]
\begin{center}
\includegraphics[scale=1]{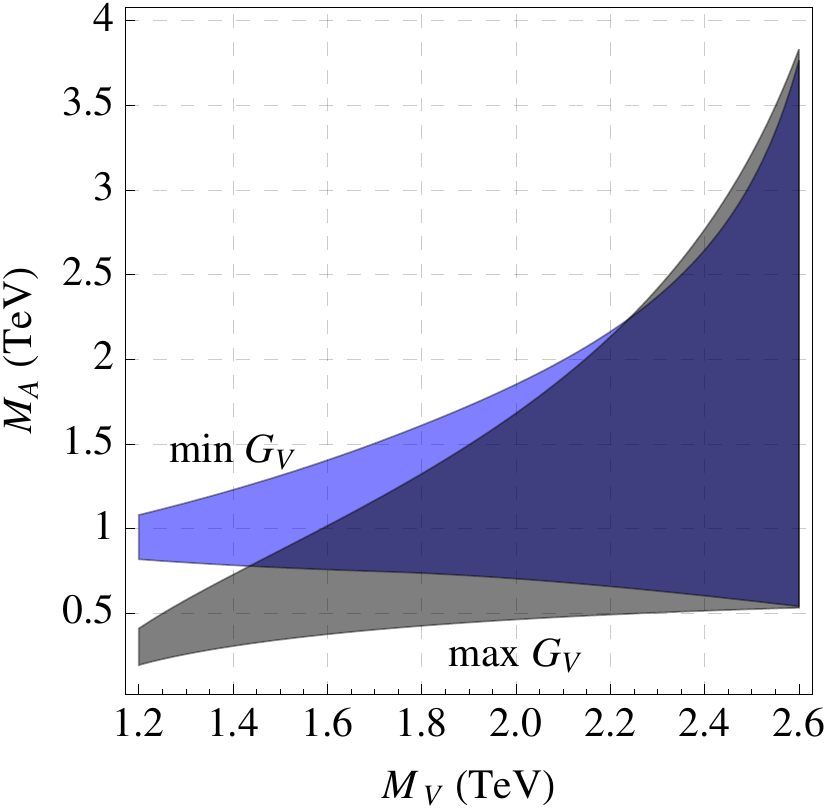}
\caption{Constraint on $M_A$, $M_V$ from the $S$ parameter; see the text for an explanation.}
\label{fig:MAMV}
\end{center}
\end{figure}

We can now compute the $S$ parameter using Eq.~\reef{eq:SVMD} as a function of $M_V$, $M_A$ (we will neglect the $O(g^2/96\pi^2)$ error term in that formula). For each $M_V$, there is a range of $M_A$ for which the computed value of the $S$ parameter gives a point in the horizontal projection of the experimental $\epsilon_1,\epsilon_3$ ellipse. 
These ranges of $M_A$ are shown in Fig.~\ref{fig:MAMV} as a function of $M_V$. The two allowed regions in that figure correspond to choosing the minimal and the maximal allowed $G_V$ for each $M_V$. For intermediate $G_V$ the allowed region changes smoothly between the shown extreme cases. We see that lower values of $M_A$ become allowed when we increase $G_V$.
This can be explained as follows. A larger $G_V$ gives a smaller $F_V$ via \reef{eq:FVGV}, which in turn produces a smaller $F_A$ from the first Weinberg sum rule. Hence smaller $M_A$ is possible without conflict with the $S$ parameter.

Demanding that the second Weinberg sum rule be also satisfied would restrict the allowed parameter space of Fig.~\ref{fig:MAMV} to a tiny region near the maximal allowed masses: $M_V\sim 2.6$ TeV, $M_A\sim 4$ TeV. As we will see below, this region is not the one preferred by the EWPT. We are led to assume that the second Weinberg does not hold. As discussed in section \ref{sec:UVtail}, this assumption is actually a necessity in models of strong EWSB based on the Conformal Technicolor idea.

Notice that we do not impose the constraint that the $S$ parameter be positive, since such an inequality has never been rigorously proved (see however \cite{Agashe} for a discussion in the context of holographic models). 

\begin{figure}[htbp]
\begin{center}
\includegraphics[scale=0.8]{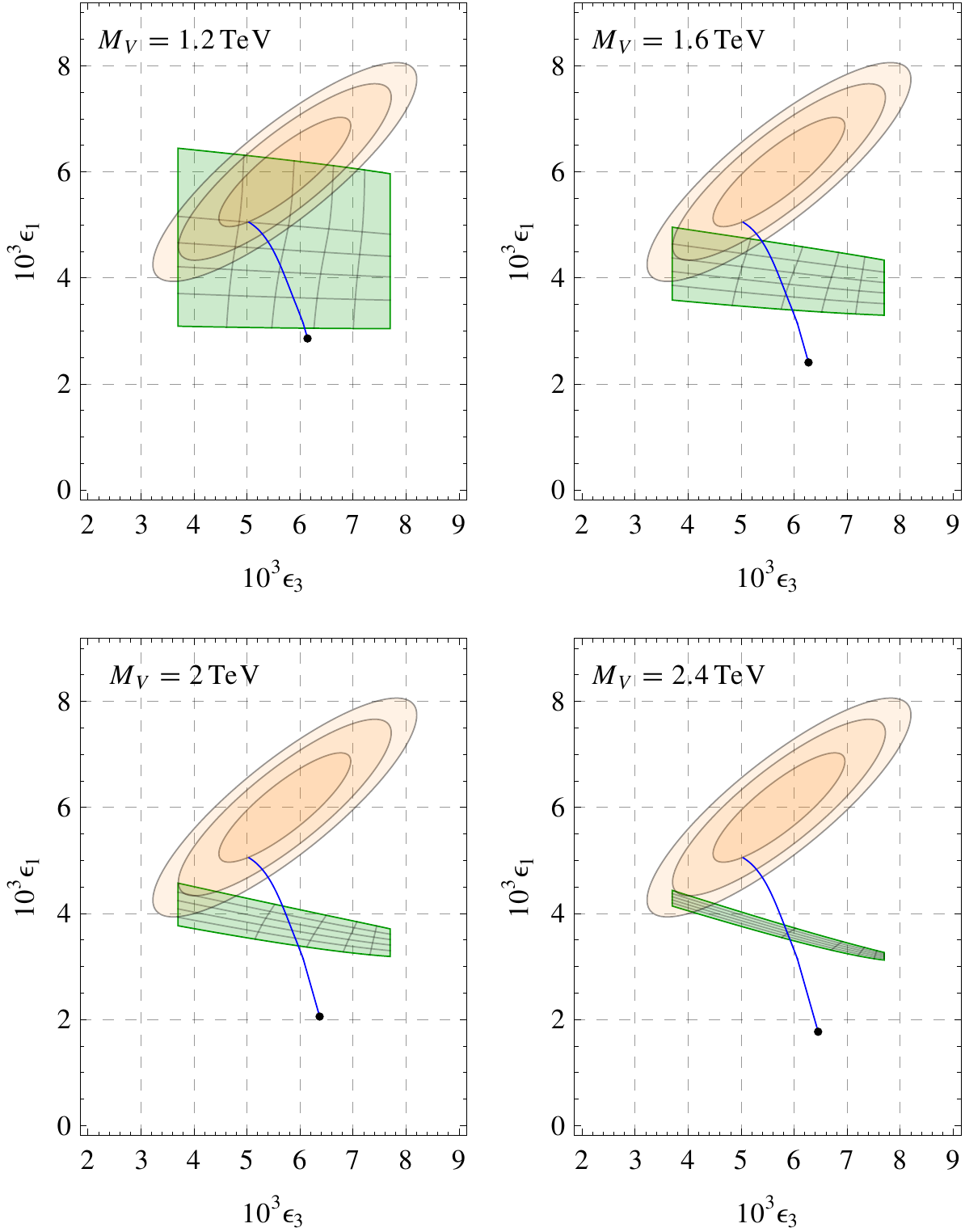}
\caption{$\lambda=-0.5$.}
\label{fig:mhalf}
\end{center}
\end{figure}

Finally, we compute the $T$ parameter as explained in section \ref{sec:TVMD}. It gets contributions from diagram (a), Eq.~\reef{eq:3a}, and diagram (b$'$) whose value is reported in Appendix \ref{sec:Tbulky}.  Figs.~\ref{fig:zero}, \ref{fig:phalf}, \ref{fig:mhalf}
then show the position of the model in the $\epsilon_3,\epsilon_1$ plane as a function of various free parameters.

In Fig.~\ref{fig:zero}, we analyze the case when only the operators \reef{eq:L2V} are added to the Lagrangian \reef{eq:LV} in order to cancel the quadratic sensitivity of the $T$ parameter to the cutoff. The couplings of these operators are thus fixed from Eq.~\reef{eq:kappaVA}. We choose four representative values $M_V=$1.2, 1.6, 2, 2.4 TeV. For each value of $M_V$, we vary $G_V$ from the minimal to maximal values allowed by Figs.~\ref{fig:GVMV}. For each $M_V,G_V$ we then vary $M_A$ from the minimal to maximal value allowed by the $S$ parameter constraint (represented in Fig.~\ref{fig:MAMV} for the extreme values of $G_V$).

The position of the model then varies within the shaded (light green) curved rectangular region. The $T$ parameter increases with increasing $G_V$, while the $S$ parameter increases with increasing $M_A$. The horizontal (vertical) curvy lines within the shaded region correspond to varying $G_V$ ($M_A$) in steps of 20\% within the allowed intervals. Finally, to guide the eye, for each $M_V$ the blue curve traces the Higgs mass dependence in the SM, terminating at $M_H=M_V$.

\begin{figure}[htbp]
\begin{center}
\includegraphics[scale=0.8]{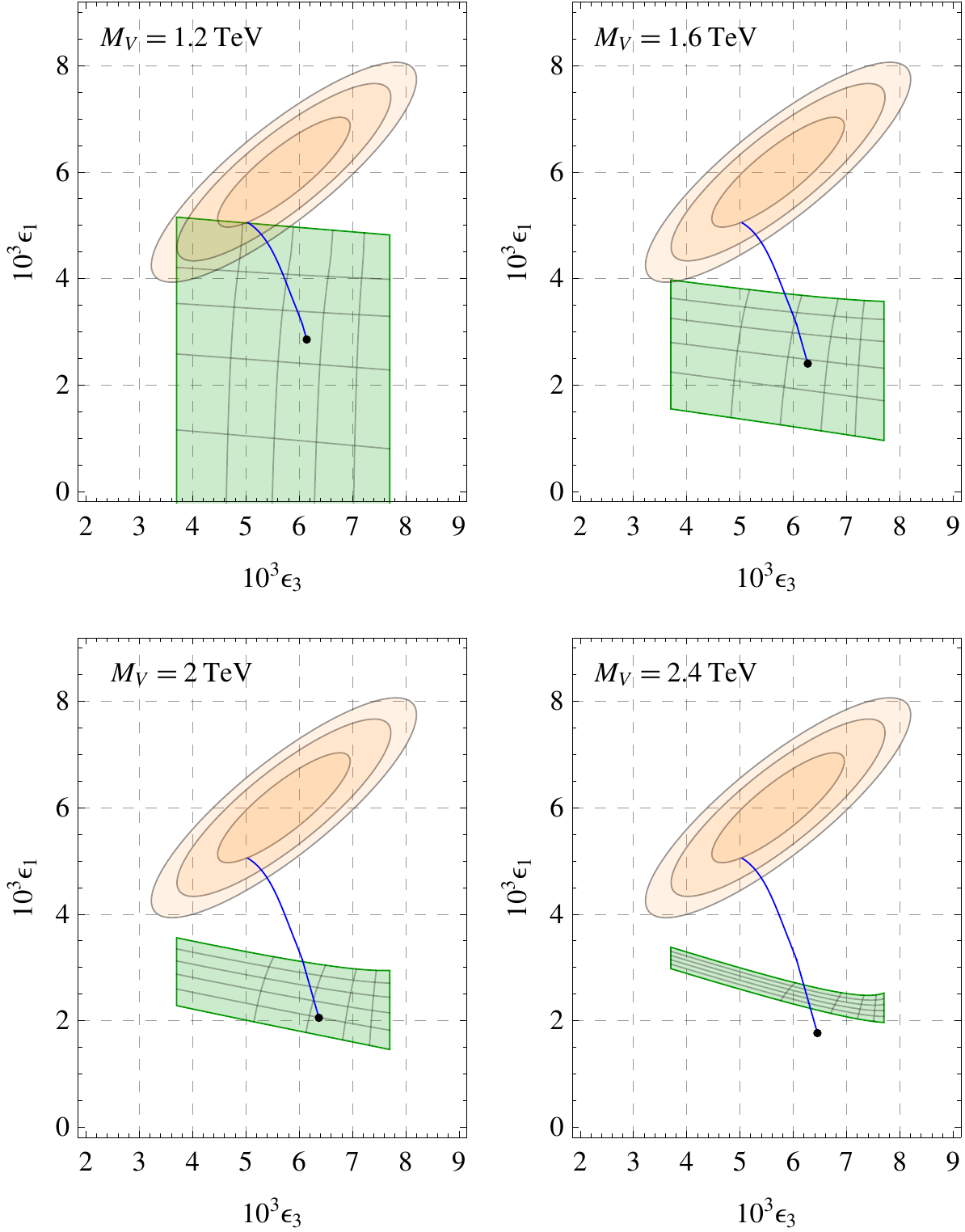}
\caption{$\lambda=0$.}
\label{fig:zero}
\end{center}
\end{figure}

In Figs.~\ref{fig:phalf} and \ref{fig:mhalf} we do the same, but adding the coupling $+i\lambda \mathscr{O}_1$ to the Lagrangian.
As mentioned in section \ref{sec:TVMD} and shown in Appendix \ref{sec:piAV}, the operator $\mathscr{O}_1$ is the only one among extra possible cubic couplings from Eq.~\reef{eq:newops} which affects the $T$ parameter. Moreover, its contribution is at most logarithmically divergent. We choose $\lambda=\pm 0.5$ in these two figures, which is of the same order of magnitude as the couplings $\kappa_{V,A}$ typically required to cancel the quadratic sensitivity of the $T$ parameter.

In all the three plots we chose the cutoff $\Lambda=2M_V$ to evaluate the logarithmically divergent terms in the $T$ parameter. We have also neglected the terms $O(M_R^2/\Lambda^2)$. We checked that these terms are numerically small.  We also checked that increasing the cutoff up to $\Lambda=3M_V$ does not change these plots significantly.

On the basis of these plots, we can now summarize the compatibility of our model with the experimental constraints. For $\lambda=0$, the agreement with the data can be obtained only in a small region of the parameter space, namely for low values of $M_V\sim 1.2$ TeV,  for $G_V$ close to the maximal values allowed by the elastic unitarity constraint (Fig.~\ref{fig:zero}).

\begin{figure}[htbp]
\begin{center}
\includegraphics[scale=0.8]{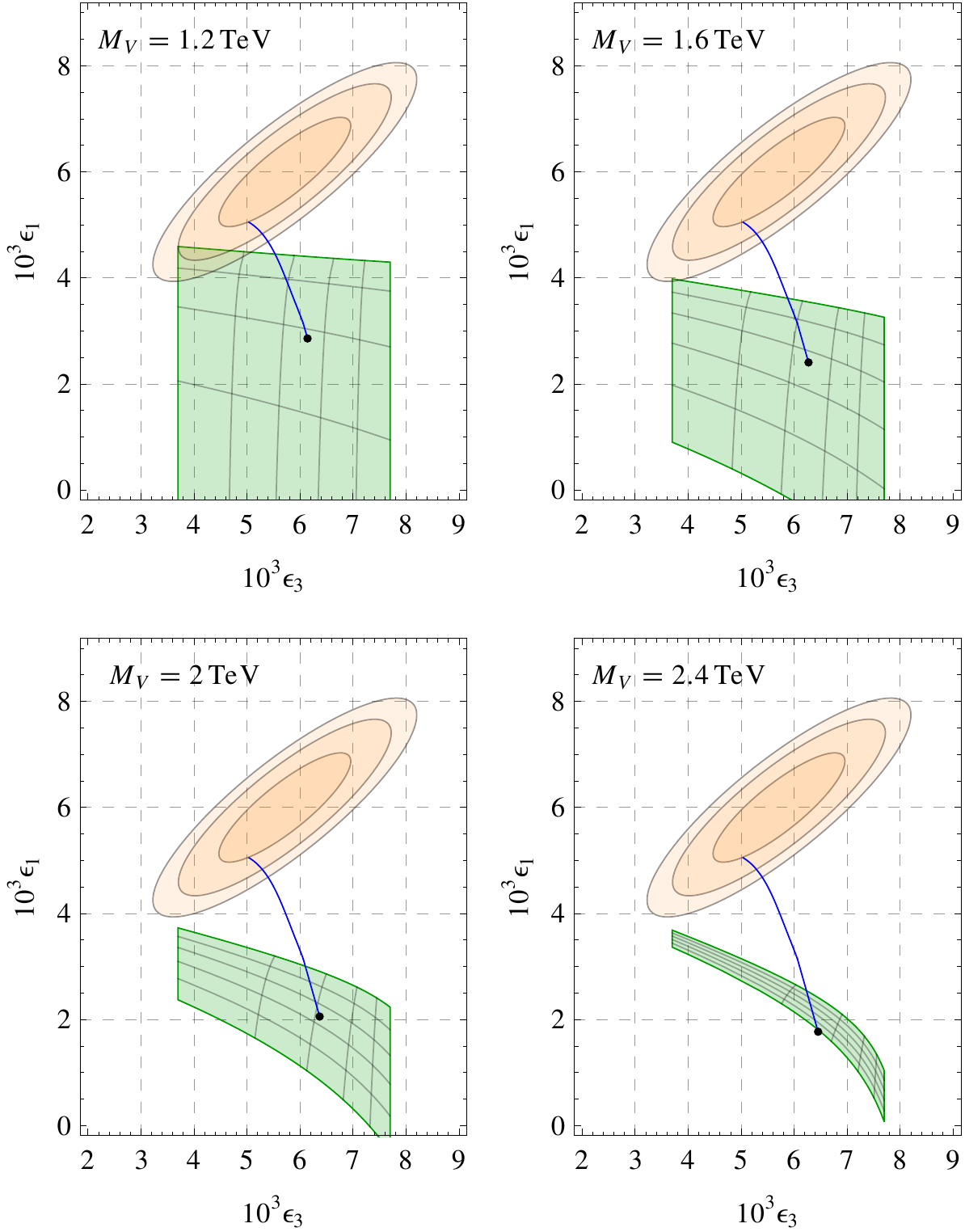}
\caption{$\lambda=+0.5$.}
\label{fig:phalf}
\end{center}
\end{figure}

Proceeding to the case when the operator $\mathscr{O}_1$ is turned on, we see that for positive $\lambda=0.5$ (Fig.~\ref{fig:phalf}) the fit is even worse that for $\lambda=0$. On the other hand, for negative $\lambda=-0.5$ the situation looks quite a bit better (Fig.~\ref{fig:mhalf}). Smaller $M_V\lesssim 1.5$ TeV are still preferred, but now the EWPT consistency holds in a reasonably wide range of $M_A$ and $G_V$. 

To summarize, we see that these models can be rendered compatible with the EWPT, provide that $V$ and $A$ are not too heavy, and especially if the extra $\pi A V$ coupling $\lambda$ is given a moderately negative value.

We would like to recall here the original idea of Ref.~\cite{us}: that after adding the operators \reef{eq:L2V} and canceling the quadratic divergence in $T$, there would remain uncanceled finite $\Delta T>0$ which could improve the electroweak fit. Indeed, our formulas for $T$ in Appendix~\ref{sec:Tbulky} contain finite terms whose origin can be traced back to the quadratic divergence \reef{eq:Tquad}. However, these terms remain subdominant in the region of parameter space allowed by all the other constraints that we impose. Basically, this happens because the mass hierarchy between $A$ and $V$ is never sufficiently large for these terms to dominate. Thus, the idea of Ref.~\cite{us} is not realized in our framework, although our basic conclusion that the resonances can provide a positive $\Delta T$ is the same.

\section{Conclusions}
\label{sec:concl}

In this paper, we have considered a generic low energy description of Higgsless scenarios of strong EWSB.
We considered one vector ($V$) and one axial ($A$) spin-1 resonance and assumed Vector Meson Dominance, i.e.~that tree level exchanges of these resonances cure the UV behavior of various formfactors, of the current two point functions, and of the elastic $W_LW_L$ scattering up to the cutoff $\Lambda \sim 2 M_V$. 

By requiring unitarity up to this scale, we fixed the value of $G_V$ in terms of $M_V$.
Assuming the good UV behavior of the pion formfactor allowed us then to fix $F_V$ in terms of $G_V$.
The first Weinberg sum rule gave then the coupling $F_A$. Anticipating the usual conflict of QCD-like scenarios with the EWPT, we assumed that the second Weinberg sum rule is not satisfied, leaving the masses unconstrained. This is also to be expected if the UV structure of the model is of Conformal Technicolor type.

We then computed the electroweak precision observables, using the Peskin-Takeuchi dispersive relation for the $S$ and a Feynman diagram approach for the $T$. 
We saw that the $T$ parameter features quadratic divergences, and
used the mechanism proposed in \cite{us} to cancel them, introducing two new cubic $\pi A V$ couplings. These quadratic divergences then leave a physical effect, as the finite terms remaining after the cancellation contain a piece of the same structure but with the cutoff replaced by $M_A$ or $M_V$. It was assumed in \cite{us} that, for this reason, keeping only those terms may be a good approximation. However, in our more restrictive framework this is not the case, since the allowed region for $M_A$ and $M_V$ does not allow large mass hierarchies.

On top of the operators of \cite{us}, we found a third cubic $\pi A V$ operator also contributing to the $T$ parameter, albeit only logarithmically.
Unfortunately we were not able to predict the corresponding coupling $\lambda$, and used an order-of-magnitude estimates to see its influence on the numerical result.
Despite this little drawback, this operator turns out to be very interesting for comparison with the data. Indeed, without it, consistency with the EWPT is possible only in a tiny corner of the parameter space where our formula for the $T$ begins to be less accurate (for $M_V \sim 1\div 1.2$ TeV and $M_A \sim 200 \div 300$ GeV). However, for negative values of $\lambda$ we can get a good fit, with values of $M_A$ less close to the electroweak scale. 

The main conclusion is thus the same as of \cite{us}: that this kind of strong EWSB scenarios can be made compatible with the EWPT using the positive $\Delta T$ from resonances. However, the inner workings of how we achieve the consistency are different.

We are sharply aware of the fact that to arrive at this conclusion we had to make a number of simplifying assumptions whose accuracy may be difficult to assess. First, we assumed that the one-loop approximation for the $T$ parameter is accurate or at least gives an idea of the expected size of the effect. Second, to reduce the number of parameters and to obtain a predictive framework, we assumed that just one generation of resonances (one vector and one axial) is important for saturating the various VMD sum rules. We hope that the noble goal of exploring the difficult world of strongly coupled models partly justifies our choice of imperfect means, for lack of better ones.

{\bf Note added.} One month after this work was posted to arXiv, the LHC experiments announced first evidence for a Higgs-boson-like particle near 125 GeV. While the working assumption of our work was that a Higgs boson does not exist, some of the lessons that we learned here will be also useful for computing electroweak precision observables in composite Higgs models, and in particular for putting on more solid grounds the estimates of $T$ and $S$ made in Ref.~\cite{extended}. 

\newpage

\begin{center} 
{\bf Acknowledgements} 
\end{center}
We are grateful to Roberto Contino and Riccardo Rattazzi for useful discussions. We thank J.~Kamenik and O.~Cata for discussions related to resolving discrepancies with their paper \cite{Kamenik} (see notes \ref{k0},\ref{k1},\ref{k2}), and to M.~Frandsen for bringing \cite{Appel} to our attention.
This work was supported in part by the European Program ``Unification in the LHC Era",
contract PITN-GA-2009-237920 (UNILHC) and by the National Science Foundation under Grant No. NSF PHY05-51164. S.~R. is grateful to KITP, Santa Barbara, for hospitality.

\appendix
\section{Experimental values of $\epsilon_1$ and $\epsilon_3$}
\label{sec:eps13exp}
The final experimental paper about the LEP and SLC precision electroweak measurements reports the following determination of the $\epsilon$ parameters (\cite{LEP}, App.E)
\begin{equation}
\begin{split}
\epsilon_1 &=+(5.4\pm1)\,10^{-3} \\
\text{\bf 2006:}\quad \epsilon_2 &=-(8.9\pm1.2)\,10^{-3} \\
\epsilon_3&=+(5.3\pm0.9)\,10^{-3} 
\end{split}
\quad\text{with correlation matrix}\quad
\rho=\left[
\begin{array}
[c]{ccc}%
1 & 0.6 & 0.86\\
0.6 & 1 & 0.4 \\
0.86 & 0.4 & 1
\end{array}\right]\,.
\label{eq:epsvalue}
\end{equation}
This determination used the preliminary measurement of the $W$ mass:
\begin{equation}
\text{\bf 2006:}\qquad M_W = 80.425\pm 0.034 \text{ GeV}\,,
\end{equation}
while the current value is \cite{EWWG}
\begin{equation}
\text{\bf 2010:}\qquad M_W = 80.399\pm 0.023 \text{ GeV}\,.
\end{equation}
This change affects the $\epsilon_2$ parameter via the term 
\begin{equation}
\epsilon_2=\frac{s_W^2}{c_W^2-s_W^2}\Delta r_\text{w}+...
\end{equation}
Here $s_W$ is the sine of the Weinberg angle in whatever definition, $s_W^2\approx 0.23, \ c_W^2=1-s_W^2$,
and $\Delta r_\text{w}$ is related to $M_W$ via 
\begin{equation}
\left(1-\frac{M_W^2}{M_Z^2}\right)\frac{M_W^2}{M_Z^2}=\frac{\pi \alpha(M_Z^2) }{\sqrt{2}G_F M_Z^2}\frac{1}{1-\Delta r_\text{w}}\,.
\end{equation}
The updated $M_W$ measurements thus leads to a shift in the central value of $\epsilon_2$, and reduces its uncertainty:
\begin{equation}
\textbf{2010:}\qquad \epsilon_2=-(8.2\pm1)\,10^{-3}\,.
\end{equation}

On the other hand, $\epsilon_{1,3}$ as well as their covariances with $\epsilon_2$ are not affected. (Recall that $\epsilon_{1,3}$ are linear combinations of $\Delta \rho$ and $\Delta k$ which are not correlated with $\Delta r_\text{w}$.) For completeness, we report the new correlation matrix:
\begin{equation}\text{\bf 2010:}\qquad\rho=\left[
\begin{array}
[c]{ccc}%
1 & 0.73 & 0.86\\
0.73 & 1 & 0.49 \\
0.86 & 0.49 & 1
\end{array}\right]\,.
\end{equation}

In most models, the $\epsilon_2$ parameter receives dominant contributions from the $W,Z$ boson loops. It is extremely weakly sensitive to the Higgs boson or other heavy particles. For example, $\epsilon_2^{\rm SM}=-(7.45\div7.35)\,10^{-3}$ for $M_H=100\, \text{GeV} \div  1\, \text{TeV}$.
For this reason, it makes sense to fix $\epsilon_2\rightarrow -7.4\times10^{-3}$, which leads to the improved determination of $\epsilon_{1,3}$ given in Eq.\ (\ref{eq:eps13exp}). A small upward shift in the central values of $\epsilon_{1,3}$ is explained by their positive correlation with $\epsilon_2$. We have checked that this determination is in very good agreement with the \emph{ab initio} fit performed recently by GFitter \cite{GFitter}.

Our final comment concerns the position of the Standard Model in the $\epsilon_1$, $\epsilon_3$ plane, Fig.~\ref{fig:e13SM}. We took $\epsilon_1^{\rm SM}=5.06\times10^{-3}$; $\epsilon_3^{\rm SM}=5\times10^{-3}$ ($M_H=111\ \text{GeV}, \ M_t=173\ \text{GeV}$), from \cite{LEP}, Table G.6, as a quick way to get a reference point. We then traced the full curve using formulas from Ref.\ \cite{Novikov} for the full $M_H$ dependence at one loop. 

\section{$T$ parameter from $\pi AV$ couplings}
\label{sec:piAV}
Here we discuss the role of the four extra operators (\ref{eq:newops}) for the computation of the $T$ parameter.

Since the only vertices that matter are those involving two resonances and one pion field, we can replace the covariant derivatives by the normal ones and expand to first order in pions. The operator $\mathcal{O}_4$ is then equivalent to  $\mathcal{O}_3$, since $\nabla_{\rho} u_{\mu}\to \partial_{\rho} \partial_{\mu}\hat{\pi}$ is at this order symmetric in $\rho,\mu$.
 
The operator $\mathcal{O}_2$ gives a contribution to the $\pi V B$, $\pi A B$ formfactor which is already proportional to $q^2$ (two derivatives acting on the pion field). 
Thus, to get a nonvanishing contribution to the $T$ parameter from the diagrams
\begin{center}
\centering
\includegraphics[scale=0.4]{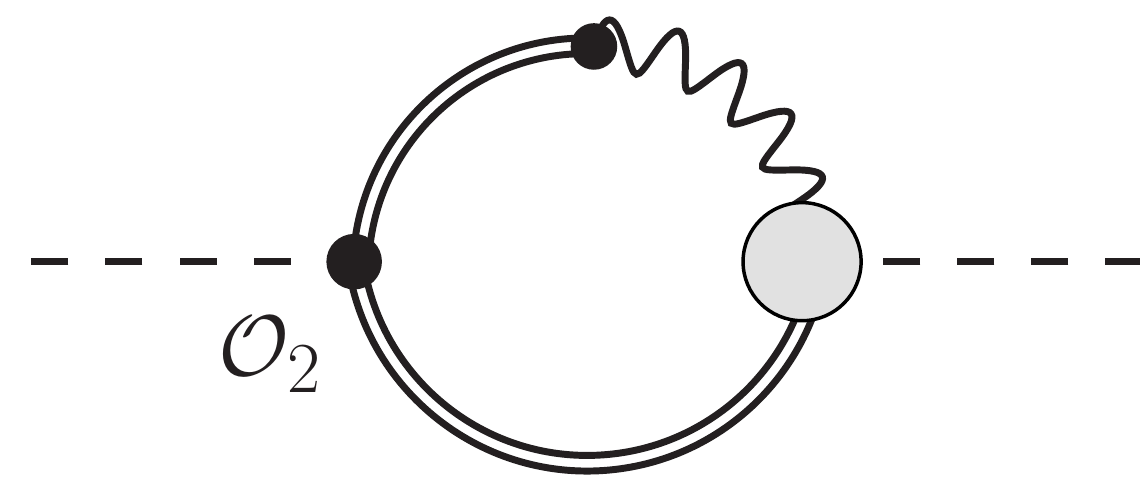}\raisebox{16pt}{\ ,}
\end{center}
we need to pick up terms in the right formfactor (blob) that are nonvanishing as $q\to0$. The only operators with no derivatives acting on the pion field are the ones corresponding to the couplings $F_V$ and $F_A$.
The only diagrams involving the $\mathcal{O}_2$ operator, that contain terms of order $q^2$ and not higher, are therefore:
\begin{center}
\begin{minipage}[b]{.25\textwidth}
\centering
\includegraphics[scale=0.4]{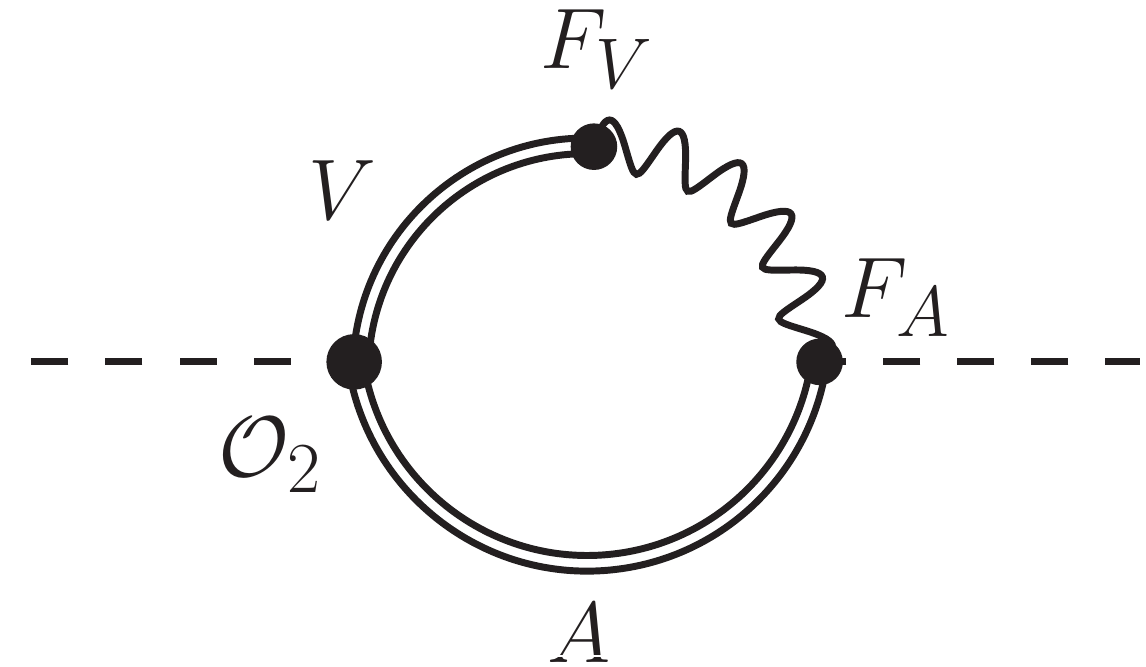}
\end{minipage} 
\raisebox{2.39em}{\qquad and\quad} 
\begin{minipage}[b]{.25\textwidth}
\centering
\includegraphics[scale=0.4]{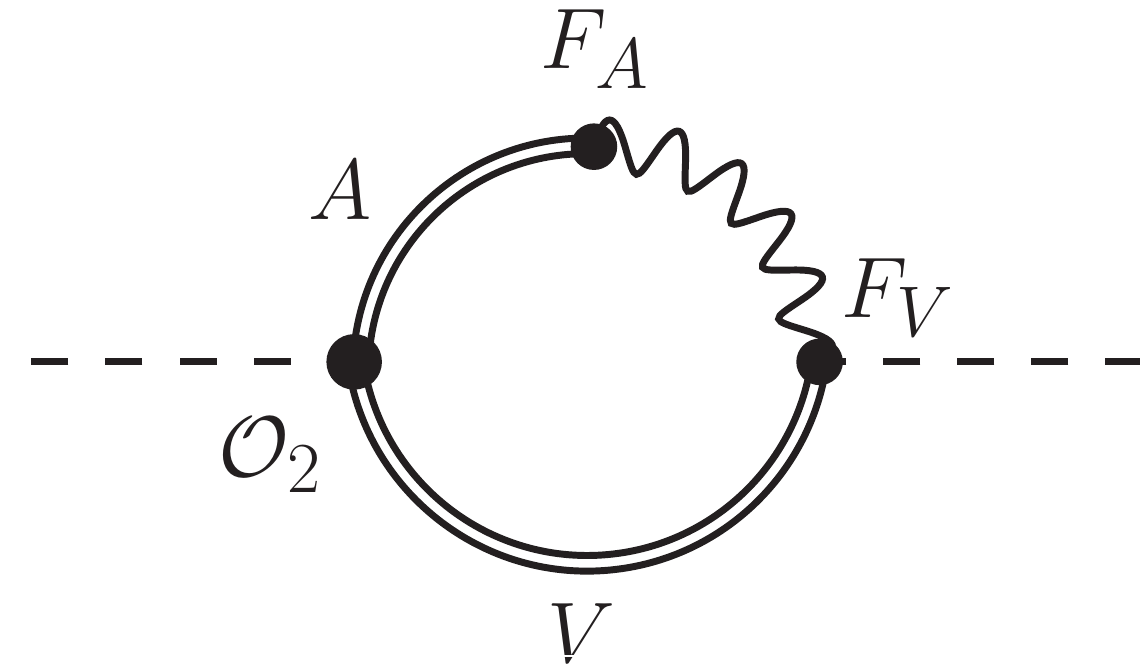} 
\end{minipage}\raisebox{2.39em}{\qquad .} 
\end{center}
But since $\mathcal{O}_2$ is antisymmetric with respect to the exchange of $A$ and $V$, these two diagrams give contributions opposite to each other. Therefore, the $\mathcal{O}_2$ operator does not contribute to the $T$ parameter. Similarly, the $\mathcal{O}_3$ does not contribute either (it is also antisymmetric in $A$ and $V$, and of order $q^2$ in the pion field).

\begin{figure}[!h]
\begin{center}
\includegraphics[scale=0.4]{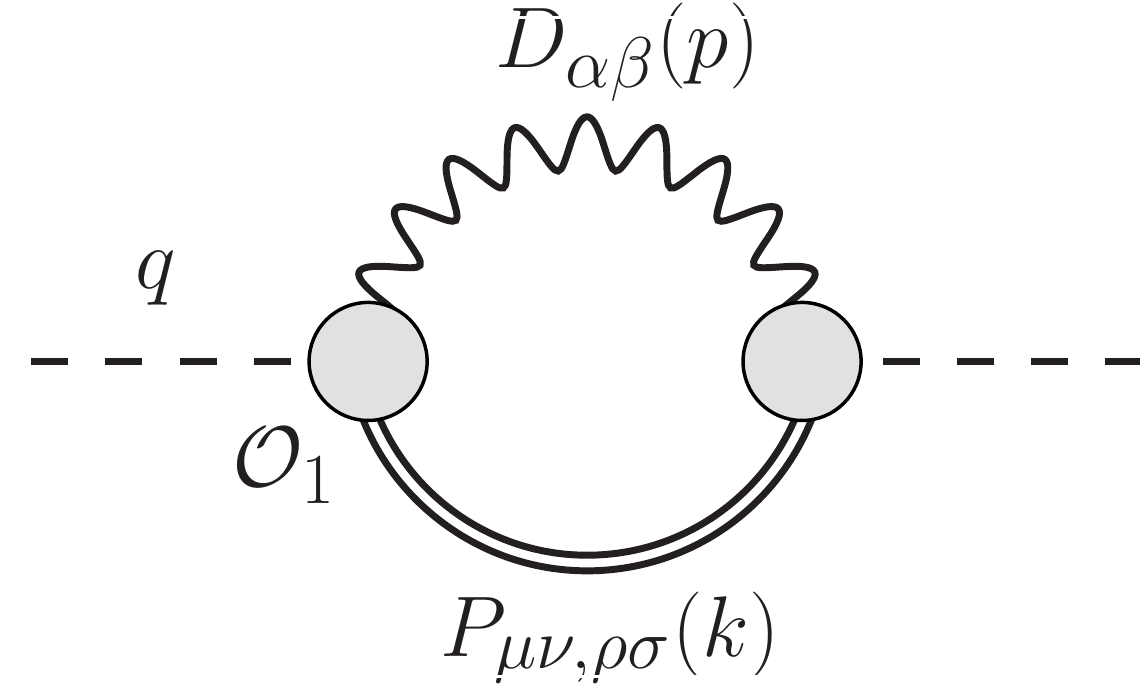}
\caption{The left blob represents the contribution to the $\pi V B$, $\pi A B$ formfactors due to $\mathcal{O}_1$. The right blob represent the full $\pi V B$, $\pi A B$ formfactors, including corrections from all the operators. $D_{\alpha\beta}$ is the $B$ propagator in the Landau gauge. The $P_{\mu\nu,\rho\sigma}$ is the $O(k^0)$ part of the resonance propagator, see Eq.~\reef{eq:Rprop1}.}
\label{fig:diago1}
\end{center} 
\end{figure}

So, we are left with $\mathcal{O}_1$. 
Let's investigate the nature of the divergent contribution to the $T$ parameter produced by this operator.
Quadratic divergences may arise only if the $O(k^0)$ term of the resonance propagator is involved; they are contained in the diagram of Fig.~\ref{fig:diago1}. 

The $\mathcal{O}_1$ contributions to $\pi VB$ and $\pi AB$ formfactors have the schematic form:
\begin{gather}
\pi VB:~\sim (p.q)g^{\alpha [\mu}p^{\nu]}, \qquad\qquad \pi AB:~\sim (k.q)g^{\alpha [\mu}p^{\nu]}\,.
\label{eq:O1ff}
\end{gather}
This is easy to understand: the $\pi$ momentum is always contracted with the momentum of $A$ (be it an outgoing leg or the resonance mixing with the $B$). The remaining Lorentz structure depends only on the $B$ momentum $p$ and must be antisymmetric in $\mu,\nu$. When we contract the  $\mathcal{O}_1$ formfactor with the $O(k^0)$ part of the resonance propagator, we can replace $p\to q$ in \reef{eq:O1ff}, because $k^{\nu}P_{\mu\nu,\rho\sigma}=0$. The LHS of the diagram in Fig.~\ref{fig:diago1} is therefore $O(q^2)$.\footnote{This would not be true for the operators in \reef{eq:L2V} which give rise to formfactors $\sim g^{\alpha [\mu}q^{\nu]}$ without extra powers of $q$. This is ultimately why \reef{eq:L2V} give rise to quadratically divergent $T$ and $\mathcal{O}_1$ not.}

As to the RHS, there are several Lorentz structures that can appear in the general formfactor. However, taking into account the antisymmetry in $\rho,\sigma$ they are all proportional to either $q$ or $k$. In the first case the diagram is $O(q^3)$ and does not contribute to the $T$ parameter, while in the second case it simply vanishes after contracting with $P_{\mu\nu,\rho\sigma}$.

Consequently, $\mathcal{O}_1$ does not give rise to quadratic divergences. However, it does give logarithmically divergent and finite contributions from the non-transverse part of the resonance propagator. We report these in Appendix \ref{sec:Tbulky}.

Finally, we remark that there are also ``double-trace" operators made out of two resonance fields and one $u_{\mu}$ invariant under the global symmetry group and parity, like
\beq 
\left\langle A^{\mu\rho} V_{\nu\rho}\right\rangle \left\langle \nabla_{\mu}u^{\nu}\right\rangle
\eeq
But since the expansions start at second order in the pion field, such operators won't contribute to the $T$ parameter.

\section{Full formulas for the $T$ parameter}
\label{sec:Tbulky}

Here we report the full formulas for the contribution of diagram (b$'$) of section \ref{sec:TVMD} to the $T$ parameter.

Diagram (b$'$) with $V$ propagating in the loop contributes ${3 g'^2}/{(256 \pi ^2  v^2)}$ times
\beq
{\footnotesize
\begin{split}
 &\frac{2 F_A^2 F_V^2+F_A^4-F_V^4}{F_V^2} \log \frac{\Lambda^2}{M_V^2}\\
 &-\frac{\xi  \left[F_A^2+F_V \left(F_V-2 G_V\right)\right] \left[(\xi -2) F_A^2+(\xi -6) F_V^2+2 \xi  G_V F_V\right]}{(\xi -1)^2 F_V^2} \log \frac{M_A^2}{M_V^2} \\
&-\frac{\xi  \left[F_A^2+F_V \left(F_V-2 G_V\right)\right] \left[F_A^2+5 F_V^2-2 G_V F_V \right]}{(\xi -1) F_V^2}+2\xi \left(F_V-2 G_V\right)^2\\
&+8\left[\frac{\lambda  F_A \left(2 \lambda  F_A F_V+F_A^2-F_V^2\right)}{F_V} \log \frac{\Lambda^2}{M_V^2}\right .\\
& + \frac{\lambda  \xi  F_A \left[-2 \lambda  (\xi -2) F_A F_V-(\xi -2) F_A^2+ (\xi +2) F_V^2-2 G_V F_V\right]}
{(\xi-1)^2 F_V} \log \frac{M_A^2}{M_V^2} \\
& \left. -\frac{\lambda  \xi  F_A \left[2 \lambda  F_A F_V+F_A^2+F_V \left(3 F_V-2 G_V\right)\right]}{(\xi -1) F_V} \right]+O(M_{V,A}^2/\Lambda^2)\,.
\end{split}
}
\label{eq:b'V}
\eeq
Here $\xi\equiv M_A^2/M_V^2$. The first three lines group terms originating from the Lagrangian \reef{eq:LV} with operators \reef{eq:L2V} added. The couplings $\kappa_{V,A}$ are fixed from Eq.~\reef{eq:kappaVA} to cancel the quadratic divergence. This explains the appearance of $F_{V,A}$ in the denominators. The second three lines group terms which appear when the coupling $+i\lambda \mathscr{O}_1$ is turned on.   

In the same notation, diagram (b$'$) with $A$ propagating in the loop contributes ${3 g'^2}/{(256 \pi ^2  v^2)}$ times
\beq
{\footnotesize
\begin{split}
 &\frac{F_A^4-2 F_V (F_V-2 G_V)F_A^2- F_V^2 (F_V-2 G_V)^2}{F_A^2}  \log \frac{\Lambda^2}{M_V^2}\\
 &-\frac{2 \xi  (\xi +2) F_A^2 F_V \left(F_V-2 G_V\right)+\left[(\xi -8) \xi +2\right] F_A^4-\xi ^2 F_V^2 \left(F_V-2 G_V\right){}^2}{(\xi -1)^2 F_A^2} \log \frac{M_A^2}{M_V^2} \\
& +\frac{\left[F_A^2+F_V \left(F_V-2 G_V\right)\right] \left[5 F_A^2+F_V \left(F_V-2 G_V\right)\right]}{(\xi -1) F_A^2} +2\xi^{-1} F_A^2\\
&+8\left[ \frac{\lambda  F_V \left[2 \lambda  F_A F_V+3 F_A^2+F_V \left(F_V-2 G_V\right)\right]}{F_A} \log \frac{\Lambda^2}{M_V^2}\right .\\
&-\frac{\lambda  \xi ^2 F_V \left[2 \lambda  F_A F_V+3 F_A^2+F_V \left(F_V-2 G_V\right)\right]}{(\xi -1)^2 F_A} \log \frac{M_A^2}{M_V^2} \\
& +\left. \frac{\lambda  F_V \left[2 \lambda  F_A F_V+3 F_A^2+F_V \left(F_V-2 G_V\right)\right]}{(\xi -1) F_A}\right]+O(M_{V,A}^2/\Lambda^2)\,.
\end{split}
}
\label{eq:b'A}
\eeq

It is instructive to compare these formulas with one partial case mentioned in the main text: setting $F_V=2G_V$, $F_A=0$, and $\lambda=0$, the logarithmically divergent term in Eq.~\reef{eq:b'V} agrees with Eq.~\reef{eq:3blog}.  

The term $\propto \xi$ which we separated in the third line of \reef{eq:b'V} is the remnant of the first quadratically divergent term 
in Eq.~\reef{eq:Tquad}. This divergence is cancelled at $\Lambda\sim M_A$. Analogously, the second quadratic divergence in Eq.~\reef{eq:Tquad} is cancelled at $\Lambda\sim M_V$ (if $M_A\ll M_V$). The remnant of this divergence is the term $\propto \xi^{-1}$ in the third line of \reef{eq:b'A}.

\end{document}